\documentclass[journal]{IEEEtran}
\usepackage[bookmarks,colorlinks]{hyperref}
\usepackage{enumitem}
\usepackage[linesnumbered,ruled,lined]{algorithm2e}

\usepackage{graphicx}
\usepackage{dcolumn}
\usepackage{bm}

\usepackage[usenames,dvipsnames]{xcolor}

\usepackage{tabularx}
\usepackage{array}

\usepackage{amsmath,amsfonts}
\usepackage{mathtools} 

\usepackage{array}
\usepackage[caption=false,font=normalsize,labelfont=sf,textfont=sf]{subfig}
\usepackage{textcomp}
\usepackage{stfloats}
\usepackage{url}
\usepackage{verbatim}
\usepackage{graphicx}
\usepackage{cite}
\usepackage{svg}
\usepackage[noend]{algpseudocode} 
\usepackage{etoolbox}
\usepackage{booktabs}   
\usepackage{array}      
\usepackage{multirow}

\usepackage{float}
\usepackage{pgfplots}
\usepackage[bookmarks,colorlinks]{hyperref}
\usepackage[linesnumbered,ruled,lined]{algorithm2e}
\usepackage{enumitem}
\usepackage{algpseudocode}
\usetikzlibrary{shapes.multipart,intersections}
\usepackage{cite}
\usepackage{amsmath,amssymb,amsfonts,amsthm,steinmetz}
\usepackage{graphicx}
\usepackage{mathrsfs}  
\usepackage{textcomp}
\usepackage{acronym}
\usepackage{xcolor}
\usepackage{upgreek,xspace}

\usepackage{tikz}
\usetikzlibrary{calc}
\makeatletter
\newcommand{\gettikzxy}[3]{%
  \tikz@scan@one@point\pgfutil@firstofone#1\relax
  \edef#2{\the\pgf@x}%
  \edef#3{\the\pgf@y}%
}
\makeatother

\usepackage[draft]{todonotes}

\usepackage{esvect}

\usepackage{times}
\usepackage{bm}
\usepackage{amsmath}
\usepackage{amssymb}
\usepackage{stmaryrd}
\usepackage{babel}
\usepackage{graphics, graphicx}
\usepackage{xcolor}
\usepackage{gensymb}
\usepackage{cite}
\usepackage{enumitem}
\usepackage{url}

\ifCLASSINFOpdf
\else
\fi
\begin{document}
%
\title{Virtual VNA 3.0: Unambiguous Scattering Matrix Estimation for \textit{Non-Reciprocal} Systems by Leveraging Tunable and Coupled Loads}
%
%
%

\author{Philipp~del~Hougne,~\IEEEmembership{Member,~IEEE}
\thanks{This work was supported in part by the ANR France 2030 program (project ANR-22-PEFT-0005), the ANR PRCI program (project ANR-22-CE93-0010), the European Union's European Regional Development Fund, and the French region of Brittany and Rennes Métropole through the contrats de plan État-Région program (projects ``SOPHIE/STIC \& Ondes'' and ``CyMoCoD'').}
\thanks{P.~del~Hougne is with Univ Rennes, CNRS, IETR - UMR 6164, F-35000 Rennes, France (e-mail: philipp.del-hougne@univ-rennes.fr).}
}

\maketitle

\begin{abstract}
We present the ``Virtual VNA 3.0'' technique for estimating the scattering matrix of a \textit{non-reciprocal}, linear, passive, time-invariant device under test (DUT) with $N$ monomodal ports using a single measurement setup involving a vector network analyzer (VNA) with only $N_\mathrm{A}<N$ ports --  thus eliminating the need for any reconnections. We partition the DUT ports into $N_\mathrm{A}$ ``accessible'' and $N_\mathrm{S}$ ``not-directly-accessible'' (NDA) ports. We connect the accessible ports to the VNA and the NDA ports to the ``virtual VNA ports'' of a VNA Extension Kit. This kit enables each NDA port to be terminated with three distinct individual loads or connected to neighboring DUT ports via coupled loads.
We derive both a closed-form and a gradient-descent method to estimate the complete scattering matrix of the non-reciprocal DUT from measurements conducted with the $N_\mathrm{A}$-port VNA under various NDA-port terminations. 
We validate both methods experimentally for $N_\mathrm{A}=N_\mathrm{S}=4$, where our DUT is a complex eight-port transmission-line network comprising circulators.
Altogether, the presented ``Virtual VNA 3.0'' technique constitutes a scalable approach to unambiguously characterize a many-port \textit{non-reciprocal} DUT with a few-port VNA (only $N_\mathrm{A}>1$ is required) -- without any tedious and error-prone manual reconnections susceptible to inaccuracies. The VNA Extension Kit requirements match those for the ``Virtual VNA 2.0'' technique that was limited to reciprocal DUTs.
\end{abstract}


\IEEEpeerreviewmaketitle

\section{Introduction}
\label{sec_introduction}

The scattering matrix of linear, passive, time-invariant systems with monomodal ports (e.g., antenna arrays, circuits) is typically measured with a vector network analyzer (VNA) in the radio-frequency, microwave and millimeter-wave regimes. To that end, each port of the device under test (DUT) is connected to a distinct port of the VNA via a monomodal transmission line (e.g., a coaxial cable). However, in the following scenarios, this conventional approach may not be viable:
\begin{enumerate}[label=(\roman*)]
    \item Some DUT ports are embedded and hence physically inaccessible, e.g., in integrated antenna arrays and circuits~\cite{zhang2019overview}, or wireless sensing applications in non-destructive testing and bioelectronics.
    \item The DUT has substantially more ports than the VNA.
    \item Connecting transmission lines to the DUT ports may significantly alter the DUT's electromagnetic properties, e.g., when characterizing miniaturized antennas~\cite{icheln1999reducing,skrivervik2001pcs,fukasawa2019investigation}.
\end{enumerate}
The common conceptual denominator of these scenarios is that the DUT has not-directly-accessible (NDA) ports, which we define as ports through which waves cannot be injected and/or received due to one or more of the aforementioned reasons. Throughout this paper, we assume that the set of the DUT's NDA ports is fixed. 
Although waves cannot be injected/received via NDA ports, it may nonetheless be feasible to terminate NDA ports with tunable loads. This raises the question whether one can unambiguously estimate the DUT's full $N \times N$ scattering matrix based on observations of how the measurable scattering at the DUT's $N_\mathrm{A}$ accessible ports depends on the terminations of the DUT's $N_\mathrm{S}$ NDA ports, where $N=N_\mathrm{A}+N_\mathrm{S}$. 

Considering a \textit{reciprocal} DUT with $N_\mathrm{S}$ NDA ports, we recently demonstrated experimentally that the DUT's full scattering matrix can be estimated free of any ambiguity, subject to the following four requirements~\cite{del2024virtual,del2024virtual2p0}:
\begin{enumerate}[label=(\arabic*)]
    \item The DUT has at least two accessible ports.~\cite{del2024virtual2p0}.
    \item Each NDA port can be terminated by three arbitrary (but distinct and known) loads~\cite{del2024virtual}. (The three available loads may differ from one NDA port to the next~\cite{del2024virtual}.)
    \item At least one accessible and at least one NDA port can be connected by an arbitrary (but known) multi-port load network (MPLN)~\cite{del2024virtual2p0}.
    \item Groups of NDA ports can be connected by arbitrary (but known) MPLNs such  that no subset of NDA ports remains for which no member can be connected via an MPLN to at least one of the other NDA ports not included in the subset~\cite{del2024virtual2p0}. A simple way to meet this requirement is to use two-port load networks (2PLN) for pairs of neighboring NDA ports~\cite{del2024virtual2p0}.
\end{enumerate}
On the one hand, we derived and validated a closed-form method requiring specific configurations of the individual loads, yielding an upper bound on the number of required measurements under low-noise conditions. On the other hand, we developed and validated a gradient-descent method that is compatible with random (opportunistic) load configurations and allows one to flexibly adapt the number of measurements to the noise level. Moreover, the gradient-descent method is compatible with intensity-only detection. We coined the ensemble of these methods ``Virtual VNA'' because the tunable loads act like ``virtual'' additional VNA ports. The MPLNs are required to eliminate sign ambiguities on off-diagonal scattering coefficients associated with NDA ports (and blockwise phase ambiguities in the case of intensity-only detection); the inclusion of requirements (ii) and (iii) was the key upgrade from the original minimal-ambiguity ``Virtual VNA''~\cite{del2024virtual} to the ambiguity-free ``Virtual VNA 2.0''~\cite{del2024virtual2p0}.

However, the ``Virtual VNA 2.0'' is still limited to \textit{reciprocal} DUTs because it assumes that the DUT's scattering matrix is symmetric. In this paper, we present and experimentally validate the generalized ``Virtual VNA 3.0'' that does not assume reciprocity such that it can be applied to \textit{non-reciprocal} DUTs. We derive a closed-form method and develop a gradient-descent method, both based on scattering parameters. The hardware requirements for the ``Virtual VNA 3.0'' coincide exactly with those listed above for the ``Virtual VNA 2.0''. We experimentally validate both ``Virtual VNA 3.0'' methods considering a DUT that is an eight-port non-reciprocal circuit (a complex transmission-line network containing circulators) in the $450-700$~MHz frequency band.

The remainder of this paper is organized as follows. 
In Sec.~\ref{sec_related_literature}, we contextualize the Virtual VNA concept with respect to related literature.
In Sec.~\ref{sec_problem_statement}, we define the measurement problem studied in this paper.
In Sec.~\ref{sec_methods}, we introduce two complementary methods to solve this measurement problem: a closed-form method (Sec.~\ref{subsec_closed_form}) and a gradient-descent method (Sec.~\ref{subsec_grad_desc}).
In Sec.~\ref{sec_exp_validation}, we experimentally validate the two methods.
We close with a brief conclusion in Sec.~\ref{sec_conclusion}.

\section{Related Literature}\label{sec_related_literature}

The ``Virtual VNA'' concept stands on the shoulders of substantial prior work on related but distinct problems. 

Decades-old ``untermination'' methods~\cite{garbacz1964determination,bauer1974embedding} are limited to DUTs with a single NDA port; variants thereof, often only targeting the retrieval of the NDA port's reflection coefficient and given substantial a priori knowledge about the DUT, have been explored in metrology and antenna characterization~\cite{garbacz1964determination,bauer1974embedding,mayhan1994technique,davidovitz1995reconstruction,pfeiffer2005recursive,pfeiffer2005equivalent,pfeiffer2005characterization,pursula2008backscattering,bories2010small,van2020verification,sahin2021noncontact,kruglov2023contactless}.

Meanwhile, ``port-reduction'' methods~\cite{lu2000port,lu2003multiport} from the early 2000s chiefly differ from the problem that we tackle in that the set of NDA ports is not fixed but varies across measurements. Because the accessible DUT ports change, the connections between accessible DUT ports and VNA ports must be changed between measurements, which is possible but potentially error-prone and tedious in scenario (ii) [DUT has more ports than VNA], but impossible in scenarios (i) [embedded DUT ports] and (iii) [DUT perturbed by transmission lines]. Moreover, no matrix-valued closed-form approach nor any gradient-descent approaches have been proposed. 

Within the realm of antenna characterization, three groups of works on multi-port backscatter
modulation techniques exist. In all three cases, the antenna system under test (ASUT) is surrounded by free space and its $N_\mathrm{S}$ ports are NDA; together with the accessible ports of $N_\mathrm{A}$ accessible auxiliary antennas, an $N$-port DUT comprising the ASUT and the auxiliary antennas can be defined.  First,~\cite{wiesbeck1998wide,monsalve2013multiport} considered scenarios with $N_\mathrm{S}=2$ and sought to retrieve the corresponding $2\times 2$ diagonal block of the the DUT's full scattering matrix, relying on a priori knowledge about the DUT and simplifying approximations. Moreover, no matrix-valued closed-form approach nor any gradient-descent approaches have been proposed. Second,~\cite{shilinkov2024antenna}\footnote{Note that while~\cite{shilinkov2024antenna} is related work, it does not predate the ``Virtual VNA'' because~\cite{shilinkov2024antenna} appeared in June 2024, after the preprint of the original ``Virtual VNA''~\cite{del2024virtual} was posted on arXiv in March 2024.} proposed a closed-form technique resembling that derived as part of the original Virtual VNA work~\cite{del2024virtual} except that it is limited to scenarios with a single accessible port and was not validated experimentally. The lack of a matrix-valued closed-form version applicable to DUTs with multiple accessible ports increases the number of required measurements and the vulnerability to noise. Third,~\cite{buck2022measuring} proposed a technique which relies on connecting in turn each port of the ASUT to a signal generator and measuring the radiation pattern for two distinct terminations with calibration-standard loads on the remaining $N_\mathrm{S}-1$ ASUT ports. This technique is specific to antenna arrays (i.e., not applicable to circuits). Moreover, the set of accessible ports changes between measurements, requiring potentially error-prone and tedious changes of the connection between DUT and signal generator in scenario (ii) [DUT has more ports than VNA] and excluding the application to scenarios (i) [embedded DUT ports] and (iii) [DUT perturbed by transmission lines].

Some of these prior works lifted ambiguities based on a priori knowledge about the DUT, e.g., leveraging the DUT's known behavior near dc~\cite{pfeiffer2005characterization} or known geometric details~\cite{shilinkov2024antenna}. Meanwhile, the importance of MPLNs to lift ambiguities without a priori knowledge was already recognized in~\cite{denicke2012application} for the estimation of parts of the scattering matrix of a reciprocal system in the context of MIMO RFID. 

In the context of optimally focusing with an array of $N_\mathrm{A}$ antennas on a load-modulated target port embedded inside an unknown medium, the DUT is the unknown medium with $N=N_\mathrm{A}+1$ ports, of which one (the target port) is NDA. Hence, this scenario assumes that the DUT has only one NDA port. Various closed-form methods (but no gradient-descent method) were proposed (see~\cite{sol2024optimal} and references therein) to perfectly focus on the target port. However, perfect focusing only requires a vector that is collinear with (but not necessarily equal to) the transmission vector from the accessible ports to the NDA port. Hence, this problem only requires an ambiguous estimation of one block of the DUT's full scattering matrix (and this block is in fact a vector because $N_\mathrm{S}=1$).

In the context of end-to-end physics-compliantly estimating the parametrization of wireless channels by a reconfigurable intelligent surface (RIS), each RIS element contains a tunable lumped element that can be understood as a lumped port terminated by a tunable load. Thus, the DUT is the radio environment (potentially comprising complicated scattering objects as well as structural scattering~\cite{king1949measurement,hansen1989relationships,hansen1990antenna} by antennas and RIS elements) and has $N=N_\mathrm{A}+N_\mathrm{S}$ ports, where $N_\mathrm{A}$ is the number of antennas and $N_\mathrm{S}$ the number of RIS elements. In this wireless communications context, only the accurate prediction of the measurable wireless channels matters. Hence, there is no need to remove ambiguities in the estimated scattering matrix of the DUT, as long as the measurable channels are accurately predicted, which reduces the above-listed four requirements to being able to terminate each NDA port with two arbitrary (but distinct, not necessarily known) individual loads. In~\cite{sol2023experimentally}, a gradient-descent method for this problem was proposed and experimentally validated, including a case restricted to intensity-only detection;\footnote{The methods in~\cite{sol2023experimentally} assume no a prior knowledge of the radio environment. If the entire RIS-parametrized radio environment is perfectly known, its full scattering matrix can of course be recovered in a single full-wave simulation by replacing the RIS elements' tunable loads with lumped ports~\cite{tapie2023systematic}, irrespective of the radio environment's complexity. When the radio environment is simply free space, a full-wave simulation of the RIS alone in combination with geometric knowledge of the radio environment is sufficient~\cite{macromodeling_ris,Naffouri2024}.} the same method was extended to so-called ``beyond-diagonal'' RIS based on their physics-compliant diagonal representation in~\cite{del2024physics}. However, these methods do not lift all ambiguities (and typical RIS-parametrized radio environments do not meet the four above-listed requirements for that). Also, so far, no corresponding closed-form method has been proposed.

Except for some ``port-reduction'' methods, to the best of our knowledge, all of the discussed related literature relies on the assumption that the DUT is reciprocal, thereby excluding by construction applications to non-reciprocal DUTs.

\section{Problem Statement}\label{sec_problem_statement}

\begin{figure*}
    \centering
    \includegraphics[width=1.5\columnwidth]{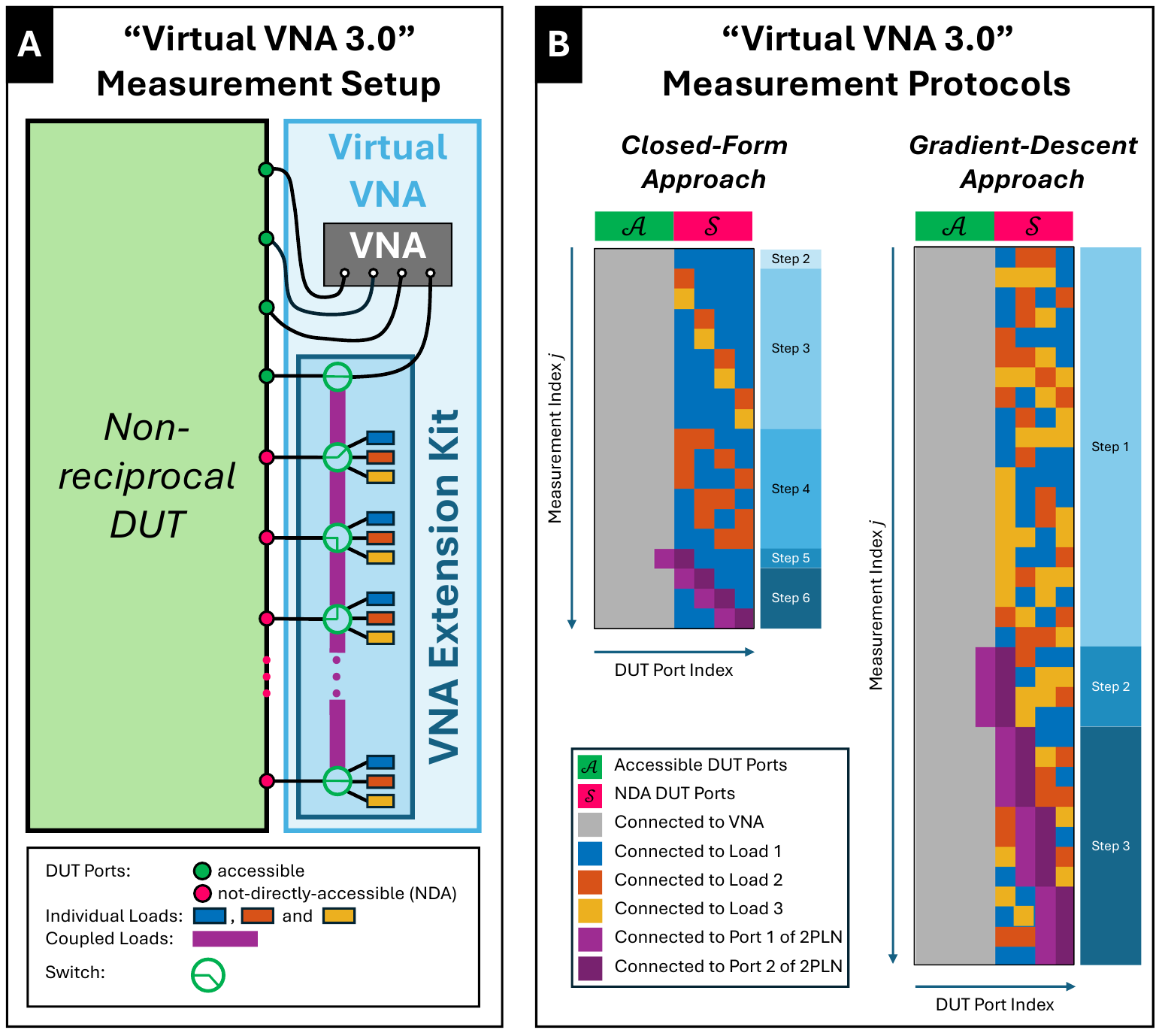}
    \caption{Overview of ``Virtual VNA 3.0'' measurement setup (A) and measurement protocols (B). The goal is to determine the scattering matrix of a \textit{non-reciprocal} $N$-port DUT with an $N_\mathrm{A}$-port VNA, where $1<N_\mathrm{A}<N$. The DUT ports are partitioned into $N_\mathrm{A}$ accessible ports that are directly connected to the VNA and $N_\mathrm{S}=N-N_\mathrm{A}$ not-directly-accessible (NDA) ports via which no waves are injected/received. The NDA ports are connected to a ``VNA Extension Kit`` that can terminate each NDA port with three distinct individual loads (blue, red, yellow), or connect pairs of neighboring NDA ports via a two-port load network (2PLN, purple). The last accessible and first NDA ports can also be connected by a 2PLN (purple). For concreteness and in line with our proof-of-principle experiments, we have chosen $N_\mathrm{A}=N_\mathrm{S}=4$ for these illustrations. Note that the ``Virtual VNA 3.0`` does \textit{not} require the individual loads to be identical at each NDA port, nor the coupled loads to be identical; these are shown as identical in our illustrations only for simplicity. Our closed-form approach (see Sec.~\ref{subsec_closed_form} for details) requires a fixed series of specific load configurations, whereas our gradient-descent approach (see Sec.~\ref{subsec_grad_desc} for details) is compatible with an arbitrary number of arbitrary load configurations.  }
    \label{Fig1}
\end{figure*}

We consider a linear, passive, time-invariant and \textit{non-reciprocal} DUT with $N$ monomodal ports. Our goal is to estimate the DUT's scattering matrix $\mathbf{S}\in \mathbb{C}^{N \times N}$ without any ambiguity; because of the DUT's non-reciprocity, $\mathbf{S}$ is \textit{not} symmetric, i.e., $\mathbf{S}\neq \mathbf{S}^\top$. We do not assume any prior knowledge about the DUT. Only $1<N_\mathrm{A}<N$ DUT ports are accessible and can be connected to an $N_\mathrm{A}$-port VNA; the remaining $N_\mathrm{S}=N-N_\mathrm{A}$ DUT ports are NDA. The sets $\mathcal{A}$ and $\mathcal{S}$ comprise, respectively, the indices of the DUT's accessible and NDA ports; these sets are fixed. No transmission line can be connected to the NDA ports to inject/receive waves, but the NDA ports are terminated by tunable loads, as illustrated in Fig.~\ref{Fig1}A. Specifically, the $i$th NDA port can be terminated by three arbitrary distinct loads with known reflection coefficients $r_i^\mathrm{A}$, $r_i^\mathrm{B}$ and $r_i^\mathrm{C}$. None of these three loads is required to coincide with a calibration standard or with those of another NDA port. Moreover, the $N_\mathrm{A}$th accessible port and the first NDA port can be connected to the first and second port, respectively, of an arbitrary known 2PLN with scattering matrix $\mathbf{S}^{\mathrm{2PLN},1} \in \mathbb{C}^{2\times 2}$. In addition, the $i$th and $(i+1)$st NDA ports can be connected to the first and second port, respectively, of an arbitrary known 2PLN with scattering matrix $\mathbf{S}^{\mathrm{2PLN},i+1} \in \mathbb{C}^{2\times 2}$. Conceptually, it does not matter whether the $N_\mathrm{S}$ 2PLNs are different from each other or not, and whether they are reciprocal or not. In practice, it is often easiest to work with similar (almost identical) and reciprocal 2PLNs. Of course,  all 2PLNs must be chosen such that the off-diagonal entries of the corresponding $2\times 2$ scattering matrix do not vanish (because otherwise the 2PLN would collapse to two individual loads and not serve its purpose). We are to date not aware of a mathematically rigorous  statement that could make the meaning of ``not vanish'' more concrete.

For a series of $M$ load configurations terminating the NDA ports, we measure the corresponding $M$ measurable $N_\mathrm{A} \times N_\mathrm{A}$ scattering matrices at the accessible ports with the $N_\mathrm{A}$-port VNA. The $j$th load configuration is characterized by the corresponding load scattering matrix $\mathbf{S}_{\mathrm{L},j}\in \mathbb{C}^{N_\mathrm{S} \times N_\mathrm{S}}$. $\mathbf{S}_{\mathrm{L},j}$ is a block-diagonal matrix containing the loads terminating the NDA ports; $\mathbf{S}_{\mathrm{L},j}$ is diagonal if all NDA ports are terminated by individual loads. The $j$th measurable scattering matrix $\hat{\mathbf{S}}_j\in\mathbb{C}^{N_\mathrm{A}\times N_\mathrm{A}}$ depends on $\mathbf{S}$ and $\mathbf{S}_{\mathrm{L},j}$ as follows~\cite{anderson_cascade_1966,ha1981solid,prod2024efficient}:
\begin{equation}
    \hat{\mathbf{S}}_j(\mathbf{S},\mathbf{S}_{\mathrm{L},j}) = \mathbf{S}_\mathcal{AA} + \mathbf{S}_\mathcal{AS} \left(  \mathbf{S}_{\mathrm{L},j}^{-1} -  \mathbf{S}_\mathcal{SS} \right)^{-1}   \mathbf{S}_\mathcal{SA},
    \label{eq_load_S}
\end{equation}
where we use the notation $\mathbf{A}_\mathcal{BC}$ to denote the block of the matrix $\mathbf{A}$ comprising rows [columns] whose indices are in the set $\mathcal{B}$ [$\mathcal{C}$]. While our closed-form approach requires a specific sequence of $M=1+3N_\mathrm{S}+N_S(N_S-1)/2$ load configurations for $N_\mathrm{A}>2$ (and one additional  measurement if $N_\mathrm{A}=2$, see Sec.~\ref{step5} below), our gradient-descent approach can be applied to arbitrarily many random load configurations (of course, the achieved accuracy depends on $M$). Illustrations of the utilized sets of load configurations for the case of $N_\mathrm{A}=N_\mathrm{S}=4$ are provided for the two approaches in Fig.~\ref{Fig1}B.

A clear embodiment of the stated problem is the characterization of non-reciprocal DUTs whose ports are physically accessible but their number exceeds the number of available VNA ports. In such an application scenario, one can declare $N_\mathrm{S}$ DUT ports as NDA and connect them to a ``VNA Extension Kit'' that can be a fairly simple printed-circuit board integrating the required ability to switch the NDA ports between different loads. This scenario is ideally suited for the purposes of this paper's proof-of-principle experiment because it allows us to directly measure the ground-truth DUT scattering matrix by directly connecting an $N$-port VNA to all DUT ports. (Of course, the ground-truth $N$-port measurement is only used for validation purposes.) In future work, we may also consider alternative embodiments of the stated problem in which the NDA ports are embedded and hence physically inaccessible. This would require the integration of tunable loads into the DUT, which was already demonstrated in experiments like~\cite{kruglov2023contactless} for a reciprocal DUT with $N_\mathrm{S}=1$.

To be clear, this problem statement and its practical embodiment only differ from those for the ``Virtual VNA 2.0''~\cite{del2024virtual2p0} regarding the fact that the DUT is non-reciprocal. In other words, the required hardware and measurement protocol are identical -- what distinguishes the ``Virtual VNA 3.0'' from its predecessor is that its post-processing does not rely on the assumption that the DUT is reciprocal, as detailed in Sec.~\ref{sec_methods}.

\section{Methods}\label{sec_methods}

In this section, we develop two complementary approaches to estimate the non-reciprocal DUT's scattering matrix without any ambiguity based on the problem statement from Sec.~\ref{sec_problem_statement}. On the one hand, we derive a closed-form approach based on scattering parameters in Sec.~\ref{subsec_closed_form}. This closed-form approach proves that the problem stated in Sec.~\ref{sec_problem_statement} can be solved, and it provides an upper bound on the number of required measurements in the low-noise regime. On the other hand, we develop a gradient-descent approach in Sec.~\ref{subsec_grad_desc} that provides flexibility by not requiring a specific set of load configurations; this flexibility enables compatibility with opportunistic load switching, and it can be used to mitigate adverse effects of measurement noise.

\subsection{Closed-Form Method}\label{subsec_closed_form}

Our closed-form method proceeds in seven major steps. Provided that an ideal matched load is available as one of the three individual loads at every NDA port, the first and last step can be skipped. Our derivation of the first and last step involves the well-established Redheffer star product~\cite{redheffer_inequalities_1959} and its less well documented inverse (which we coin the inverse Redheffer star product). We now explain these two operations. Consider two linear, passive, time-invariant scattering systems U and V with $N_\mathrm{U}$ and $N_\mathrm{V}$ ports, respectively. The corresponding scattering matrices are $\mathbf{S}^\mathrm{U}\in\mathbb{C}^{N_\mathrm{U} \times N_\mathrm{U}}$ and $\mathbf{S}^\mathrm{V}\in\mathbb{C}^{N_\mathrm{V} \times N_\mathrm{V}}$. $N_\mathrm{C}<N_\mathrm{U}$ ports from U are connected (without loss, reflection or delay) to $N_\mathrm{C}<N_\mathrm{V}$ ports from V. The ports of U [V] are hence partitioned into a set $\mathcal{C}_\mathrm{U}$ [$\mathcal{C}_\mathrm{V}$] containing the ports that are connected to V [U] and a set $\mathcal{N}_\mathrm{U}$ [$\mathcal{N}_\mathrm{V}$] containing the remaining free ports. Importantly, some ports of both U and V remain free after the connection.
The resulting connected system, referred to as UV, has $\eta = (N_\mathrm{U}-N_\mathrm{C})+(N_\mathrm{V}-N_\mathrm{C})$ free ports and is characterized by the scattering matrix $\mathbf{S}^\mathrm{UV}\in\mathbb{C}^{\eta \times \eta}$. 
Given $\mathbf{S}^\mathrm{U}$, $\mathbf{S}^\mathrm{V}$, $\mathcal{C}_\mathrm{U}$ and $\mathcal{C}_\mathrm{V}$, we can evaluate $\mathbf{S}^\mathrm{UV}$ via the (forward) Redheffer star product:
\newcommand{\U}{\ensuremath{\mathrm{U}}}
\newcommand{\V}{\ensuremath{\mathrm{V}}}
\newcommand{\iC}{\ensuremath{\mathcal{C}}}
\newcommand{\iN}{\ensuremath{\mathcal{N}}}
\newcommand{\mS}{\ensuremath{\mathbf{S}}}
\newcommand{\mI}{\ensuremath{\mathbf{I}}}
\begin{equation}
    {\mS}^{\U\V} = \mS^\U \underset{\iC_\U,\iC_\V}{\star} \mS^\V.
\label{eq:redheffer_star}
\end{equation}
Given $\mathbf{S}^\mathrm{UV}$, $\mathbf{S}^\mathrm{V}$, $\mathcal{C}_\mathrm{U}$ and $\mathcal{C}_\mathrm{V}$, we can evaluate $\mathbf{S}^\mathrm{U}$ via the \textit{inverse} Redheffer star product:
\begin{equation}
    {\mS}^{\U} = \mS^{\U\V} \overset{-1}{\underset{\iC_\U,\iC_\V}{\star}} \mS^\V.
\label{eq:inverse_redheffer_star}
\end{equation}
The definitions of the forward and inverse Redheffer star products are provided in the Appendix.

We formulate our closed-form method in terms of scattering parameters which avoids conversions between impedance and scattering parameters that were necessary in earlier Virtual VNA versions~\cite{del2024virtual,del2024virtual2p0}; however, the major contribution of this subsection is that the DUT is not assumed to be reciprocal.

\begin{figure}
    \centering
    \includegraphics[width=\columnwidth]{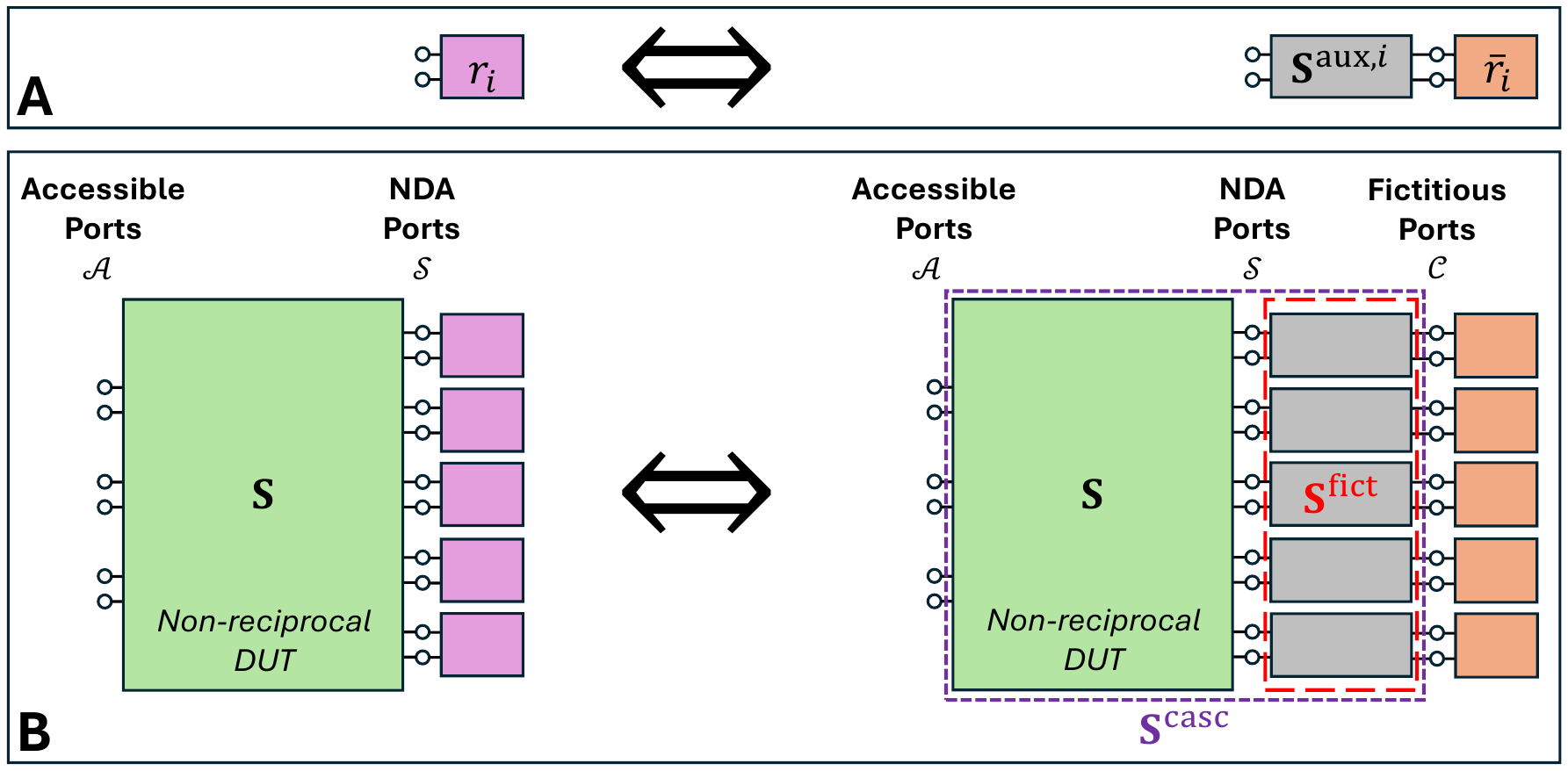}
    \caption{(A) Representation of an arbitrary one-port system with reflection coefficient $r_i$ as an equivalent cascade of a two-port system with scattering matrix $\mathbf{S}^{\mathrm{aux},i}$ and a different one-port system with reflection coefficient $\bar{r}_i$. Each port is represented as a pair of two terminals. (B) Application of (A) to the individual terminations of all NDA ports of the non-reciprocal DUT. The collection of all auxiliary two-port systems is a fictitious network with scattering matrix $\mathbf{S}^\mathrm{fict}$ defined in Eq.~(\ref{eq_6}). The cascade of the DUT with the fictitious network yields a system with scattering matrix $\mathbf{S}^\mathrm{casc}$ defined in Eq.~(\ref{eq_5}).}
    \label{Fig2}
\end{figure}

In the following, we explain the seven major steps of our closed-form method. We denote our estimate of the DUT's scattering matrix by $\tilde{\mathbf{S}}$.

\subsubsection{Define a fictitious network to pretend matched loads are available}

For steps 2-6, we would like to assume that each NDA port can be terminated by a matched load. However, this is usually not the case in practice. Hence, we select $r_i^\mathrm{A}$ as reference load for the $i$th NDA port. As illustrated in Fig.~\ref{Fig2}A, we can think of any one-port system with reflection coefficient $r_i$ as an equivalent cascade of a two-port system with scattering matrix $\mathbf{S}^{\mathrm{aux},i}$ and a different one-port system with reflection coefficient $\Bar{r}_i$, where -- analogous to Eq.~(\ref{eq_load_S}) -- we have
\begin{equation}
    r_i = \mathbf{S}^{\mathrm{aux},i}_{11} + \frac{\mathbf{S}^{\mathrm{aux},i}_{12}\mathbf{S}^{\mathrm{aux},i}_{21}}{\frac{1}{\Bar{r}_i}-\mathbf{S}^{\mathrm{aux},i}_{22}}.
    \label{eq_fict}
\end{equation}
Inspection of Eq.~(\ref{eq_fict}) reveals that we must choose $\mathbf{S}^{\mathrm{aux},i}_{11} = r_i^\mathrm{A}$ so that $r_i$ equals $r_i^\mathrm{A}$ when $\Bar{r}_i$ is a matched load. We can randomly choose complex-valued non-zero values for the other three entries of $\mathbf{S}^{\mathrm{aux},i}$; for simplicity, we impose $\mathbf{S}^{\mathrm{aux},i}_{12}=\mathbf{S}^{\mathrm{aux},i}_{21}$.  
Having fixed $\mathbf{S}^{\mathrm{aux},i}$, we can solve Eq.~(\ref{eq_fict}) for $\Bar{r}_i$ and determine the reflection coefficient $\Bar{r}_i$ for the cases in which $r_i$ equals $r_i^\mathrm{B}$ or $r_i^\mathrm{C}$; we denote these by $\Bar{r}_i^\mathrm{B}$ and $\Bar{r}_i^\mathrm{C}$. We repeat this procedure for each NDA port. 
Then, as shown in Fig.~\ref{Fig1}B, we summarize all auxiliary two-port systems in a fictitious network characterized by 
\begin{equation}\label{eq_6}
     \mathbf{S}^\mathrm{fict} = \mathrm{blockdiag}(\mathbf{S}^{\mathrm{aux},i}) \in \mathbb{C}^{2N_\mathrm{S} \times 2 N_\mathrm{S}}.
\end{equation}
The cascade of the DUT with this fictitious network is characterized by 
\begin{equation}\label{eq_5}
        {\mS}^{\mathrm{casc}} = \mS \underset{\mathcal{S},\bar{\mathcal{S}}}{\star} \mS^\mathrm{fict},
\end{equation}
where $\bar{\mathcal{S}}$ denotes the set of port indices of the fictitious network that are connected to the DUT. 
At this point, we can formulate an equivalent version of our main problem in which we seek to retrieve $\mathbf{S}^\mathrm{casc}\in\mathbb{C}^{N\times N}$ [instead of $\mathbf{S}$] and can terminate the $i$th NDA port of the cascade [instead of the DUT] by a matched load or $\Bar{r}_i^\mathrm{B}$ or $\Bar{r}_i^\mathrm{C}$ [instead of $r_i^\mathrm{A}$ or $r_i^\mathrm{B}$ or $r_i^\mathrm{C}$]. 

In steps 2-6, we derive our estimate $\tilde{\mS}^{\mathrm{casc}}$ of  ${\mS}^{\mathrm{casc}}$, and in step 7 we recover $\tilde{\mathbf{S}}$ from $\tilde{\mS}^{\mathrm{casc}}$ based on our knowledge of $ \mathbf{S}^\mathrm{fict}$.
Note that if a matched load is available at every NDA port in practice, then $\tilde{\mathbf{S}}$ and $\tilde{\mathbf{S}}^\mathrm{casc}$ are identical, implying that steps 1 and 7 can be skipped.

\subsubsection{Reference configuration: All NDA ports terminated by reference loads} In step 2, we determine the $\mathcal{AA}$ block of $\tilde{\mathbf{S}}^\mathrm{casc}$.
We terminate each NDA port with its reference load $r_i^\mathrm{A}$, implying
\begin{equation}
        \mathbf{S}_{\mathrm{L},1} = \mathrm{diag}\left(\left[r_1^\mathrm{A},r_2^\mathrm{A},\dots,r_{N_\mathrm{S}}^\mathrm{A}\right]\right).
\end{equation}
The corresponding load scattering matrix at the fictitious ports is
\begin{equation}
        \bar{\mathbf{S}}_{\mathrm{L},1} = \mathrm{diag}\left(\left[0,0,\dots,0\right]\right).
\end{equation}
By inspection of Eq.~(\ref{eq_load_S}), we identify the measurable scattering matrix $\hat{\mathbf{S}}_1$ at the accessible DUT ports as the sought-after $\mathcal{AA}$ block of $\tilde{\mathbf{S}}^\mathrm{casc}$:
\begin{equation}
     \tilde{\mathbf{S}}^\mathrm{casc}_\mathcal{AA} \triangleq \hat{\mathbf{S}}_1 .
\end{equation}

\subsubsection{Reference configuration except for one NDA port's individual load}
In step 3, we determine the diagonal entries of the $\mathcal{SS}$ block of $\tilde{\mathbf{S}}^\mathrm{casc}$, and we determine the blocks $\mathcal{AS}$ and $\mathcal{SA}$ of $\tilde{\mathbf{S}}^\mathrm{casc}$ up to row-wise and column-wise scaling factors that we fix in steps 5 and 6. 

Let us maintain all loads in the reference configuration except for the $i$th NDA port which we terminate with $r_i^\mathrm{B}$. The corresponding load scattering matrix at the fictitious ports is  
\begin{equation}
        \bar{\mathbf{S}}_{\mathrm{L},2i} = \mathrm{diag}\left(\left[0,\dots, \bar{r}_i^\mathrm{B},\dots,0\right]\right),
\end{equation}
and it follows from Eq.~(\ref{eq_load_S}) that
\begin{equation}
    \hat{\mathbf{S}}_{2i} = \mathbf{S}^\mathrm{casc}_\mathcal{AA} + \mathbf{S}^\mathrm{casc}_{\mathcal{A}\mathcal{S}_i} \left( \frac{1}{\bar{r}_i^\mathrm{B}} - \sigma_i \right)^{-1} \mathbf{S}^\mathrm{casc}_{\mathcal{S}_i\mathcal{A}},
\end{equation}
where $\mathcal{S}_i$ denotes the $i$th entry of the set $\mathcal{S}$ and $\sigma_i = \mathbf{S}^\mathrm{casc}_{\mathcal{S}_i\mathcal{S}_i}$. We define
\begin{equation}
    \Delta\mathbf{S}_{2i} = \hat{\mathbf{S}}_{2i} -  \hat{\mathbf{S}}_{1} = \mathbf{S}^\mathrm{casc}_{\mathcal{A}\mathcal{S}_i} \left( \frac{1}{\bar{r}_i^\mathrm{B}} - \sigma_i \right)^{-1} \mathbf{S}^\mathrm{casc}_{\mathcal{S}_i\mathcal{A}}, 
\end{equation}
which must be a rank-one matrix because $\mathbf{S}^\mathrm{casc}_{\mathcal{A}\mathcal{S}_i}$ and $\mathbf{S}^\mathrm{casc}_{\mathcal{S}_i\mathcal{A}}$ are vectors. Consequently, $\Delta\mathbf{S}_{2i}$ only has one non-zero singular value, and the corresponding left [right] right singular vector must be collinear with $\mathbf{S}^\mathrm{casc}_{\mathcal{A}\mathcal{S}_i}$ [$\left( \mathbf{S}^\mathrm{casc}_{\mathcal{A}\mathcal{S}_i} \right)^\star$]. 

Next, we switch the $i$th NDA port termination from $r_i^\mathrm{B}$ to $r_i^\mathrm{C}$, while maintaining all others in their reference configuration, corresponding to
\begin{subequations}
    \begin{equation}
        \bar{\mathbf{S}}_{\mathrm{L},2i+1} = \mathrm{diag}\left(\left[0,\dots, \bar{r}_i^\mathrm{C},\dots,0\right]\right),
    \end{equation}
    \begin{equation}
    \hat{\mathbf{S}}_{2i+1} = \mathbf{S}^\mathrm{casc}_\mathcal{AA} + \mathbf{S}^\mathrm{casc}_{\mathcal{A}\mathcal{S}_i} \left( \frac{1}{\bar{r}_i^\mathrm{C}} - \sigma_i \right)^{-1} \mathbf{S}^\mathrm{casc}_{\mathcal{S}_i\mathcal{A}}.
    \end{equation}
\end{subequations}

Given the measurements $\hat{\mathbf{S}}_1$, $\hat{\mathbf{S}}_{2i}$ and $\hat{\mathbf{S}}_{2i+1}$, our first goal is now to obtain the most robust estimate of a vector that is collinear with $\mathbf{S}^\mathrm{casc}_{\mathcal{A}\mathcal{S}_i}$, and similarly for $\mathbf{S}^\mathrm{casc}_{\mathcal{S}_i\mathcal{A}}$. We denote by $\mathbf{u}_i^\mathrm{AB}$, $\mathbf{u}_i^\mathrm{AC}$, and $\mathbf{u}_i^\mathrm{BC}$ the first left singular vector of $\hat{\mathbf{S}}_{2i} -  \hat{\mathbf{S}}_1$, $\hat{\mathbf{S}}_{2i+1} - \hat{\mathbf{S}}_1$ and $\hat{\mathbf{S}}_{2i+1} -  \hat{\mathbf{S}}_{2i}$, respectively. Similarly, we denote by $\mathbf{v}_i^\mathrm{AB}$, $\mathbf{v}_i^\mathrm{AC}$, and $\mathbf{v}_i^\mathrm{BC}$ the conjugates of the corresponding first right singular vectors. Then, we stack them: $\mathbf{U}_i=[\mathbf{u}_i^\mathrm{AB},\mathbf{u}_i^\mathrm{AC},\mathbf{u}_i^\mathrm{BC}]$ and $\mathbf{V}_i=[\mathbf{v}_i^\mathrm{AB},\mathbf{v}_i^\mathrm{AC},\mathbf{v}_i^\mathrm{BC}]$. Finally, we define $\mathbf{u}_i$ and $\mathbf{v}_i$ as the first left singular vector of $\mathbf{U}_i$ and $\mathbf{V}_i$, respectively. $\mathbf{u}_i$ and $\mathbf{v}_i$ are related to $\mathbf{S}^\mathrm{casc}_{\mathcal{A}\mathcal{S}_i}$ and $\mathbf{S}^\mathrm{casc}_{\mathcal{S}_i\mathcal{A}}$ as follows:
\begin{equation}\label{eq_14}
    \begin{split}
        \mathbf{S}^\mathrm{casc}_{\mathcal{A}\mathcal{S}_i} = \alpha_i \mathbf{u}_i,\\
        \mathbf{S}^\mathrm{casc}_{\mathcal{S}_i\mathcal{A}} = \beta_i \mathbf{v}_i,
    \end{split}
\end{equation}
where $\alpha_i$ and $\beta_i$ are so-far unknown complex-valued scalars. Key differences at this stage in comparison to the case of a reciprocal DUT are that generally $\mathbf{v}_i\neq\mathbf{u}_i^\top$ and $\alpha_i \neq \beta_i$. 

Having fixed $\mathbf{u}_i$ and $\mathbf{v}_i$, we would like to determine $\alpha_i$, $\beta_i$ and $\sigma_i$. However, at this stage, we can only determine $\gamma_i=\alpha_i\beta_i$ and $\sigma_i$. To that end, we first determine the complex-valued scalar $k_i^\mathrm{AB}$ such that $\hat{\mathbf{S}}_{2i}-\hat{\mathbf{S}}_{1} = k_i^\mathrm{AB} \mathbf{u}_i \mathbf{v}_i^\top$, and similarly for $k_i^\mathrm{AC}$. Then, after a few algebraic manipulations similar to those detailed in Appendix~A of~\cite{del2024virtual}, we find
\begin{subequations}
    \begin{equation}\label{eq_sigma_casc_ii}
        \sigma_i = \frac{k^\mathrm{AB}_i(\bar{r}_i^\mathrm{C}-\bar{r}_i^\mathrm{A}) - k^\mathrm{AC}_i(\bar{r}_i^\mathrm{B}-\bar{r}_i^\mathrm{A})}{k_i^\mathrm{AB}(\bar{r}_i^\mathrm{C}-\bar{r}_i^\mathrm{A})\bar{r}_i^\mathrm{B}- k_i^\mathrm{AC}( \bar{r}_i^\mathrm{B}-\bar{r}_i^\mathrm{A}) \bar{r}_i^\mathrm{C}},
    \end{equation}
    \begin{equation}\label{eq_gamma_casc_i}
        \gamma_i = \frac{
  k_i^\mathrm{AB}k_i^\mathrm{AC}\bigl(k_i^\mathrm{AB} - k_i^\mathrm{AC}\bigr)
  \bigl(\bar{r}_i^\mathrm{B} - \bar{r}_i^\mathrm{A}\bigr)
  \bigl(\bar{r}_i^\mathrm{C} - \bar{r}_i^\mathrm{A}\bigr)
  \bigl(\bar{r}_i^\mathrm{B} - \bar{r}_i^\mathrm{C}\bigr)
}{
  \Bigl[
    k_i^\mathrm{AB}\,\bar{r}_i^\mathrm{B}\,\bigl(\bar{r}_i^\mathrm{C} - \bar{r}_i^\mathrm{A}\bigr)
    \;-\;
    k_i^\mathrm{AC}\,\bar{r}_i^\mathrm{C}\,\bigl(\bar{r}_i^\mathrm{B} - \bar{r}_i^\mathrm{A}\bigr)
  \Bigr]^2
}.
    \end{equation}
\end{subequations}

We have now unambiguously retrieved the diagonal entry of $\tilde{\mathbf{S}}^\mathrm{casc}$ corresponding to the $i$th NDA port: 
\begin{equation}
    \tilde{\mathbf{S}}^\mathrm{casc}_{\mathcal{S}_i\mathcal{S}_i} \triangleq \sigma_i.
\end{equation}
Moreover, up to two distinct scaling factors for which we only know the product, we also retrieved the transmission vector from the accessible ports to the $i$th NDA port and the transmission vector from the $i$th NDA port to the accessible ports. We make the following particular (but arbitrary) choice regarding the scaling factor in our estimated cascaded scattering matrix (we fix the ambiguity in steps 5 and 6):
\begin{subequations}\label{eq_19}
    \begin{equation}
        \tilde{\mathbf{S}}^\mathrm{casc}_{\mathcal{A}\mathcal{S}_i} \triangleq \gamma_i \mathbf{u}_i,
    \end{equation}
        \begin{equation}
        \tilde{\mathbf{S}}^\mathrm{casc}_{\mathcal{S}_i\mathcal{A}} \triangleq \mathbf{v}_i.
    \end{equation}
\end{subequations}

We repeat the procedure of this step 3  for each NDA port in turn.

\subsubsection{Reference configuration except for two NDA ports' individual loads}

In step 4, we determine the off-diagonal entries of the $\mathcal{SS}$ block of $\tilde{\mathbf{S}}^\mathrm{casc}$ up to ambiguities resulting from those in step 3 that we fix in steps 5 and 6.

We now maintain all loads in the reference configuration except for the $i$th and $j$th NDA ports which we terminate with $r_i^\mathrm{B}$ and $r_j^\mathrm{B}$, respectively, with $j>i$.
The corresponding load scattering matrix at the fictitious ports has two non-zero diagonal entries:  
\begin{equation}
        \bar{\mathbf{S}}_{\mathrm{L},m(i,j)} = \mathrm{diag}\left(\left[0,\dots, \bar{r}_i^\mathrm{B},\dots,\bar{r}_j^\mathrm{B},\dots,0\right]\right),
\end{equation}
where $m(i,j)=1 + 2N_{\mathrm{S}} + \sum_{k=1}^{i-1} \bigl(2N_{\mathrm{S}} - k\bigr) + \bigl(j - i\bigr)$. It follows from Eq.~(\ref{eq_load_S}) that
\begin{equation}
    \hat{\mathbf{S}}_{m(i,j)} = \mathbf{S}^\mathrm{casc}_\mathcal{AA} + \mathbf{S}^\mathrm{casc}_{\mathcal{A}\mathcal{S}_{ij}} \left( \bar{\mathbf{S}}_{\mathrm{L},m(i,j)}^{-1} - \mathbf{S}^\mathrm{casc}_{\mathcal{S}_{ij}\mathcal{S}_{ij}} \right)^{-1} \mathbf{S}^\mathrm{casc}_{\mathcal{S}_{ij}\mathcal{A}},
\end{equation}
where $\mathcal{S}_{ij}$ denotes the set containing the $i$th and $j$th entries of the set $\mathcal{S}$. The diagonal entries of $\mathbf{S}^\mathrm{casc}_{\mathcal{S}_{ij}\mathcal{S}_{ij}}$ are $\sigma_i$ and $\sigma_j$; for ease of notation, we refer to the off-diagonal entries as $\kappa_{ij}$ and $\kappa_{ji}$ in the following. Note that generally $\kappa_{ij}\neq \kappa_{ji}$ due to the DUT's non-reciprocity. We define
\begin{equation}\label{eq_18}
\begin{split}
    \Delta\mathbf{S}_{m(i,j)} &= \hat{\mathbf{S}}_{m(i,j)} -  \hat{\mathbf{S}}_1 \\ &= \mathbf{S}^\mathrm{casc}_{\mathcal{A}\mathcal{S}_{ij}} \left( \! \begin{bmatrix}  1/\bar{r}_i^\mathrm{B}  &  0 \\ 0 & 1/\bar{r}_j^\mathrm{B}    \\ \end{bmatrix}^{-1} \! \! - \! \begin{bmatrix}  \sigma_i  &  \kappa_{ij} \\ \kappa_{ji} & \sigma_j    \\ \end{bmatrix}\!  \right)^{-1} \! \mathbf{S}^\mathrm{casc}_{\mathcal{S}_{ij}\mathcal{A}}, 
    \end{split}
\end{equation}
which must be a rank-two matrix by inspection. To mitigate measurement noise, one can replace $\Delta\mathbf{S}_{m(i,j)}$ in the following with a version of $\Delta\mathbf{S}_{m(i,j)}$  that is truncated to rank two.  

Next, we evaluate
\begin{equation}
    \mathbf{Q}_{ij} = \left( \tilde{\mathbf{S}}^\mathrm{casc}_{\mathcal{A}\mathcal{S}_{ij}} \right)^+ \Delta\mathbf{S}_{m(i,j)}  \left(  \tilde{\mathbf{S}}^\mathrm{casc}_{\mathcal{S}_{ij}\mathcal{A}} \right)^+,
\end{equation}
where $\mathbf{A}^+$ denotes  the Moore–Penrose pseudo-inverse of a matrix $\mathbf{A}$.
Meanwhile, the $2\times 2$ matrix inversions in Eq.~(\ref{eq_18}) can be evaluated in closed form. We would now like to compare the entries of $\mathbf{Q}_{ij}$ to the analytically inverted matrix from Eq.~(\ref{eq_18}) to identify $\kappa_{ij}$ and $\kappa_{ji}$. However, due to the ambiguities in $\mathbf{Q}_{ij}$ carried forward from the unknown scaling factor in Eq.~(\ref{eq_19}), we can only find ambiguous versions $\eta_{ij}$ and $\eta_{ji}$ of $\kappa_{ij}$ and $\kappa_{ji}$, respectively, at this stage. Based on a few algebraic manipulations similar to those detailed in Appendix~B of~\cite{del2024virtual}, we define the $(i,j)$th and $(j,i)$th off-diagonal entries of the $\mathcal{SS}$ block of $\tilde{\mathbf{S}}^\mathrm{casc}$ as follows:
\begin{subequations}\label{eq_21}
    \begin{equation}
        \tilde{\mathbf{S}}^\mathrm{casc}_{\mathcal{S}_i\mathcal{S}_j} \triangleq \frac{[\mathbf{Q}_{ij}]_{12}}{\chi},
    \end{equation}
        \begin{equation}
        \tilde{\mathbf{S}}^\mathrm{casc}_{\mathcal{S}_j\mathcal{S}_i} \triangleq \frac{[\mathbf{Q}_{ij}]_{21}}{\chi},
    \end{equation}
\end{subequations}
where
\begin{equation}\label{eq_22}
    \chi = \left[\left(\frac{1}{\bar{r}_i^\mathrm{B}} - \sigma_i\right) \left(\frac{1}{\bar{r}_j^\mathrm{B}} - \sigma_j\right) - \eta_{ij} \eta_{ji}\right]^{-1}
\end{equation}
and
    \begin{equation}\label{eq_23}
    \begin{split}
        \eta_{ij} \eta_{ji} &= \left(\frac{1}{\bar{r}_j^\mathrm{B}} - \sigma_j\right) \left({\frac{1}{\bar{r}_i^\mathrm{B}}} - \sigma_i\right) - \frac{1}{[\mathbf{Q}_{ij}]_{11}} \left(\frac{1}{\bar{r}_j^\mathrm{B}} - \sigma_j\right),
 \\ &= \left(\frac{1}{\bar{r}_j^\mathrm{B}} - \sigma_j\right) \left({\frac{1}{\bar{r}_i^\mathrm{B}}} - \sigma_i\right) - \frac{1}{[\mathbf{Q}_{ij}]_{22}} \left(\frac{1}{\bar{r}_i^\mathrm{B}} - \sigma_i\right).
 \end{split}
    \end{equation}
For improved robustness, we take the average of the two values of $\eta_{ij} \eta_{ji}$ obtained from Eq.~(\ref{eq_23}), inject that into Eq.~(\ref{eq_22}) to determine $\chi$, and finally inject that into Eq.~(\ref{eq_21}).

Instead of the described pseudo-inverse based procedure, the same result can also be obtained by solving multiple quadratic equations explicitly and identifying the common solution. However, the pseudo-inverse based approach presented here appears more compact; it does require $N_\mathrm{A}>1$ but this is automatically satisfied by our first requirement ``at least two accessible DUT ports'' mentioned in the introduction.
We recommend to use as many accessible DUT ports as possible to improve the robustness to noise; note that our closed-form method requires the same set of load configurations and data analysis for any $N_\mathrm{A}>1$. 

We repeat the procedure of this step 4 for each pair $(i,j)$ of NDA ports in turn, with $j>i$.

\subsubsection{Reference configuration except for a 2PLN connecting an accessible port and an NDA port} \label{step5}

Before detailing step 5, let us summarize our progress so far. We have unambiguously estimated the $\mathcal{AA}$ block and the diagonal entries of the $\mathcal{SS}$ block of $\mathbf{S}^\mathrm{casc}$. Moreover, we have identified the remainder of $\mathbf{S}^\mathrm{casc}$ up to some correlated column-wise and line-wise ambiguities. Our current $\tilde{\mathbf{S}}^\mathrm{casc}$ is of the following form:
\begin{equation}\label{eq_2555}
    \tilde{\mathbf{S}}^\mathrm{casc} = \begin{bmatrix}  \mathbf{S}^\mathrm{casc}_\mathcal{AA} & \epsilon_1 \mathbf{S}^\mathrm{casc}_{\mathcal{A}\mathcal{S}_1} &  \dots &  \epsilon_{N_\mathrm{S}} \mathbf{S}^\mathrm{casc}_{\mathcal{A}\mathcal{S}_{N_\mathrm{S}}}  \\  \epsilon_1^{-1} \mathbf{S}^\mathrm{casc}_{\mathcal{S}_1\mathcal{A}} & \mathbf{S}^\mathrm{casc}_{\mathcal{S}_1\mathcal{S}_1}  &  \dots &  \epsilon_{1}^{-1}\epsilon_{N_\mathrm{S}}\mathbf{S}^\mathrm{casc}_{\mathcal{S}_1\mathcal{S}_{N_\mathrm{S}}}  \\  \vdots
 & \vdots     &  \ddots &  \vdots\\  \epsilon_{N_\mathrm{S}}^{-1} \mathbf{S}^\mathrm{casc}_{\mathcal{S}_{N_\mathrm{S}}\mathcal{A}} & \epsilon_{N_\mathrm{S}}^{-1}\epsilon_1 \mathbf{S}^\mathrm{casc}_{\mathcal{S}_1\mathcal{S}_{N_\mathrm{S}}} &   \dots &  \mathbf{S}^\mathrm{casc}_{\mathcal{S}_{N_\mathrm{S}}\mathcal{S}_{N_\mathrm{S}}}\\ \end{bmatrix},
\end{equation}
where $\epsilon_i$ are complex-valued scalars that remain to be determined. In Eq.~(\ref{eq_19}) of step 3 we made a particular (but arbitrary) choice that implies that $\epsilon_i $ is in fact $\beta_i$, but we do not know $\beta_i$ so far. Note also that $\eta_{ij} = \epsilon_i^{-1}\epsilon_j\kappa_{ij}$ and  $\eta_{ji} = \epsilon_j^{-1}\epsilon_i\kappa_{ji}$. In the simpler case of a reciprocal DUT, the knowledge that $\mathbf{S}^\mathrm{casc}$ is symmetric would allow us to narrow $\epsilon_i$ down to $\pm 1$~\cite{del2024virtual,del2024virtual2p0}.

In this step 5, our goal is to identify $\epsilon_1$ to lift the ambiguities on the row and column associated with the first NDA port. To that end, we connect the first 2PLN to the last accessible and first NDA port, maintaining all other NDA port terminations in their reference configuration. The motivation for using this configuration is that it treats the last accessible port like an NDA port but there is no ambiguity on the row and column of $\mathbf{S}^\mathrm{casc}$ associated with it. Based on Eq.~(\ref{eq_load_S}), we thus measure \begin{equation}\label{eq_25}
\begin{split}
    \hat{\mathbf{S}}_{n} &= \mathbf{S}^\mathrm{casc}_{\mathcal{A}^\prime\mathcal{A}^\prime} + \mathbf{S}^\mathrm{casc}_{\mathcal{A}^\prime\mathcal{X}} \left( (\bar{\mathbf{S}}^{\mathrm{2PLN},1})^{-1} - \mathbf{S}^\mathrm{casc}_{\mathcal{X}\mathcal{X}} \right)^{-1} \mathbf{S}^\mathrm{casc}_{\mathcal{X}\mathcal{A}^\prime}, 
    \end{split}
\end{equation}
where $n=2+2N_\mathrm{S}+N_\mathrm{S}(N_\mathrm{S}-1)/2$, $\mathcal{A}^\prime = \mathcal{A} \setminus \mathcal{A}_{N_\mathrm{A}}$ and $\mathcal{X} = \mathcal{A}_{N_\mathrm{A}} \cup \ \mathcal{S}_1$. Note that whereas the dimensions of all scattering matrices measured in steps 1 to 4 were $N_\mathrm{A}\times N_\mathrm{A}$, the dimensions of $\hat{\mathbf{S}}_{n}$ in the present step 5 are $(N_\mathrm{A}-1)\times (N_\mathrm{A}-1)$, implying a requirement for $N_\mathrm{A}>1$, as mentioned. $\bar{\mathbf{S}}^{\mathrm{2PLN},1}$ is the 2PLN attached to the fictitious ports from $\mathcal{X}$ and generally different from ${\mathbf{S}}^{\mathrm{2PLN},1}$. Indeed, 
\begin{equation}
    {\mathbf{S}}^{\mathrm{2PLN},1} = \bar{\mathbf{S}}^{\mathrm{2PLN},1} \underset{2,1}{\star} \left( \mathbf{P} \mathbf{S}^\mathrm{aux,1} \right),
\end{equation}
where $\mathbf{P} = \left[\begin{smallmatrix} 0 & 1\\1 & 0 \end{smallmatrix}\right]$ is a permutation matrix. Hence, we determine $\bar{\mathbf{S}}^{\mathrm{2PLN},1}$ as follows:
\begin{equation}\label{eq_288}
    \bar{\mathbf{S}}^{\mathrm{2PLN},1} = \mathbf{S}^{\mathrm{2PLN},1} \overset{-1}{\underset{2,1}{\star}} \left( \mathbf{P} \mathbf{S}^\mathrm{aux,1} \right).
\end{equation}

At this stage, the only unknown in Eq.~(\ref{eq_25}) is $\epsilon_1$. Although an arguably simpler pseudo-inverse based approach (sharing some similarities with step 4) exists for $N_\mathrm{A}>2$ whose validity we confirmed numerically, we found that the following quadratic-equation based approach is more robust under real conditions. It consists in solving the following matrix-valued quadratic equation for $\epsilon_1$:
\begin{equation}\label{eq_39}
    \mathbf{0} = \mathbf{X} \epsilon_1^2 + \mathbf{Y} \epsilon_1 + \mathbf{Z},
\end{equation}
where
\begin{subequations}\label{eq_40}
\begin{equation}
    \mathbf{X} = - b_n \left(  \tilde{\mathbf{S}}^\mathrm{casc}_{\mathcal{A}^\prime\mathcal{S}_1}\tilde{\mathbf{S}}^\mathrm{casc}_{\mathcal{A}_{N_\mathrm{A}}\mathcal{A}^\prime}   + \tilde{\mathbf{S}}^\mathrm{casc}_{\mathcal{A}_{N_\mathrm{A}}\mathcal{S}_1} \left(\hat{\mathbf{S}}_n - \tilde{\mathbf{S}}^\mathrm{casc}_{\mathcal{A}^\prime\mathcal{A}^\prime}   \right) \right), 
\end{equation}
\begin{equation}
    \mathbf{Y} = \mathbf{B}-\mathbf{A} \left(\hat{\mathbf{S}}_n - \tilde{\mathbf{S}}^\mathrm{casc}_{\mathcal{A}^\prime\mathcal{A}^\prime}   \right),
\end{equation}
\begin{equation}
    \mathbf{Z} =  -b_n\left(\tilde{\mathbf{S}}^\mathrm{casc}_{\mathcal{A}^\prime\mathcal{A}_{N_\mathrm{A}}}\tilde{\mathbf{S}}^\mathrm{casc}_{\mathcal{S}_1\mathcal{A}^\prime}+\tilde{\mathbf{S}}^\mathrm{casc}_{\mathcal{S}_1\mathcal{A}_{N_\mathrm{A}}}\left(\hat{\mathbf{S}}_n - \tilde{\mathbf{S}}^\mathrm{casc}_{\mathcal{A}^\prime\mathcal{A}^\prime}   \right)\right).
\end{equation}
\end{subequations}
In Eq.~(\ref{eq_40}), we used the following auxiliary variables for notational ease:
\begin{subequations}
\begin{equation}\label{eq_41a}
    \begin{bmatrix} a_n & b_n \\ b_n & c_n \end{bmatrix} = \left( \bar{\mathbf{S}}^{\mathrm{2PLN},1}\right)^{-1},
\end{equation}
\begin{equation}
        \mathbf{A} = (a_n-\tilde{\mathbf{S}}^\mathrm{casc}_{\mathcal{A}_{N_\mathrm{A}}\mathcal{A}_{N_\mathrm{A}}})(c_n-\tilde{\mathbf{S}}^\mathrm{casc}_{\mathcal{S}_1\mathcal{S}_1}) - b_n^2 - \tilde{\mathbf{S}}^\mathrm{casc}_{\mathcal{A}_{N_\mathrm{A}}\mathcal{S}_1}\tilde{\mathbf{S}}^\mathrm{casc}_{\mathcal{S}_1\mathcal{A}_{N_\mathrm{A}}},
    \end{equation}
    \begin{equation}
    \begin{split}
        \mathbf{B} = \tilde{\mathbf{S}}^\mathrm{casc}_{\mathcal{A}^\prime\mathcal{A}_{N_\mathrm{A}}}(c_n-\tilde{\mathbf{S}}^\mathrm{casc}_{\mathcal{S}_1\mathcal{S}_1})\tilde{\mathbf{S}}^\mathrm{casc}_{\mathcal{A}_{N_\mathrm{A}}\mathcal{A}^\prime} + \tilde{\mathbf{S}}^\mathrm{casc}_{\mathcal{A}_{N_\mathrm{A}}\mathcal{S}_1}\tilde{\mathbf{S}}^\mathrm{casc}_{\mathcal{A}^\prime\mathcal{A}_{N_\mathrm{A}}}\tilde{\mathbf{S}}^\mathrm{casc}_{\mathcal{S}_1\mathcal{A}^\prime} \\+ \tilde{\mathbf{S}}^\mathrm{casc}_{\mathcal{S}_1\mathcal{A}_{N_\mathrm{A}}}\tilde{\mathbf{S}}^\mathrm{casc}_{\mathcal{A}^\prime\mathcal{S}_1}\tilde{\mathbf{S}}^\mathrm{casc}_{\mathcal{A}_{N_\mathrm{A}}\mathcal{A}^\prime} + \tilde{\mathbf{S}}^\mathrm{casc}_{\mathcal{A}^\prime\mathcal{S}_1}(a_n-\tilde{\mathbf{S}}^\mathrm{casc}_{\mathcal{A}_{N_\mathrm{A}}\mathcal{A}_{N_\mathrm{A}}})\tilde{\mathbf{S}}^\mathrm{casc}_{\mathcal{S}_1\mathcal{A}^\prime}.
        \end{split}
    \end{equation}
\end{subequations}
Recall regarding Eq.~(\ref{eq_41a}) that $\bar{\mathbf{S}}^{\mathrm{2PLN},1}$ is a symmetric matrix because of how it is defined in Eq.~(\ref{eq_288}) based on the symmetric matrices ${\mathbf{S}}^{\mathrm{2PLN},1}$ and $\mathbf{S}^\mathrm{aux,1}$. The fact that the latter two are symmetric was our choice and is independent from the fact that the DUT is non-reciprocal.

Each scalar-valued quadratic equation yields two solutions for $\epsilon_1$ of which we expect one to be correct. Altogether, we obtain $2(N_\mathrm{A}-1)^2$ solutions for $\epsilon_1$ from the $(N_\mathrm{A}-1)^2$ scalar-valued quadratic equations defined in Eq.~(\ref{eq_39}), of which we expect half to be correct and hence identical (under ideal noise-free conditions) and the other half to generally be different from the first half and mutually different. Hence, we use a standard k-Means clustering algorithm to identify $(N_\mathrm{A}-1)^2+1$ clusters, of which we expect one cluster to contain significantly more solutions than all others. Under ideal noise-free conditions, one cluster should contain the $(N_\mathrm{A}-1)^2$ identical correct solutions and the other clusters should each contain one distinct incorrect solution. We hence define our final estimate of $\epsilon_1$ as the average of the solutions contained within the largest cluster. We can now lift one row-wise and column-wise ambiguity, and hence update our estimate of $\mathbf{S}^\mathrm{casc}$ accordingly:
\begin{subequations}
    \begin{equation}
        \tilde{\mathbf{S}}^\mathrm{casc}_{\mathcal{T} \mathcal{S}_1} \leftarrow \epsilon_1^{-1} \tilde{\mathbf{S}}^\mathrm{casc}_{\mathcal{T} \mathcal{S}_1},
    \end{equation}
    \begin{equation}
        \tilde{\mathbf{S}}^\mathrm{casc}_{ \mathcal{S}_1 \mathcal{T}} \leftarrow \epsilon_1 \tilde{\mathbf{S}}^\mathrm{casc}_{ \mathcal{S}_1\mathcal{T}},
    \end{equation}
\end{subequations}
where $\mathcal{T} = \mathcal{A} \cup \mathcal{S}$.

In the special case of $N_\mathrm{A}=2$, the matrices $\mathbf{X}$, $\mathbf{Y}$ and $\mathbf{Z}$ are scalars, meaning that we have only one quadratic equation for $\epsilon_1$. Hence, Eq.~(\ref{eq_39}) yields two possible solutions of which one is the correct one but we do not know which one it is. A possible work-around to identify the correct solution in this special case of $N_\mathrm{A}=2$ consists in replacing the utilized 2PLN by a different one, and repeating the entire procedure of step 5, which would yield another quadratic equation with two solutions, of which again one is the correct one. Then, the solution that is common to both quadratic equations will be the sought-after one. Technically, we would thus have four solutions which we could divide into three clusters using a k-Means algorithm, and then define our final solution $\epsilon_1$ as the mean of the cluster containing two solutions. To summarize, in the special case of $N_\mathrm{A}=2$, one additional measurement would be required for step 5.\footnote{Note that in the case of a DUT that is known to be reciprocal, the case of $N_\mathrm{A}=2$ would not require an additional measurement in step 5 because only a binary decision between $\epsilon_1$ being equal to $+1$ or $-1$ would be required~\cite{del2024virtual2p0}.}

\subsubsection{Reference configuration except for a 2PLN connecting two NDA ports}
In step 6, we lift all remaining ambiguities by determining the remaining $\epsilon_i$ (for $1<i\leq N_\mathrm{S}$) with a procedure very similar to step 5. The basic idea is to determine $\epsilon_i$ by connecting a 2PLN to the NDA ports indexed $i$ and $j$, where $j$ is chosen such that it is an NDA port for which $\epsilon_j$ has been lifted previously. To start, we know $\epsilon_1$ from step 5. With every additional measurement, we identify one new $\epsilon_i$. Step 6 hence requires $N_\mathrm{S}-1$ measurements. The measured scattering matrices are of dimensions $N_\mathrm{A}\times N_\mathrm{A}$ (only step 5 was an exception because one accessible port was not used to inject/receive waves therein).  For the sake of conciseness, we do not print the mathematical details of step 6 because they are directly analogous to those provided for step 5. At the end of step 6, we have achieved an ambiguity-free estimate of $\mathbf{S}^\mathrm{casc}$, i.e., $\tilde{\mathbf{S}}^\mathrm{casc}$ equals $\mathbf{S}^\mathrm{casc}$ under ideal noise-free conditions.

\subsubsection{Remove the fictitious network}
In step 7, we recover $\tilde{\mathbf{S}}$ given $\tilde{\mS}^{\mathrm{casc}}$ from step 6 and knowledge of our chosen ${\mS}^{\mathrm{fict}}$ from step 1 by inverting Eq.~(\ref{eq_5}):
\begin{equation}\label{eq_33}
    \tilde{\mS} = \tilde{\mS}^\mathrm{casc} \overset{-1}{\underset{\mathcal{S},\bar{\mathcal{S}}}{\star}} \mS^\mathrm{fict}.
\end{equation}
Under ideal noise-free conditions, $\tilde{\mS}$ coincides with $\mS$. As already mentioned, if an ideal matched load is available as reference load at each NDA port, then $\tilde{\mS} = \tilde{\mS}^\mathrm{casc}$, and steps 1 and 7 are redundant.

\subsection{Gradient-Descent Method}\label{subsec_grad_desc}

In the previous subsection, we derived a closed-form approach that proves that it is possible to recover the \textit{non-reciprocal} DUT's scattering matrix free of any ambiguity with a generalized Virtual VNA approach, and that yields an upper bound on the number of required measurements under ideal noise-free conditions. In this subsection, we establish a complementary gradient-descent approach that can be applied to measurements obtained with an arbitrary number of arbitrary load configurations. In particular, this property of the gradient-descent approach provides the flexibility to adjust the number of measurements to the measurement noise.

In the following, we explain the three major steps of our gradient-descent method. We denote our estimate of the DUT's scattering matrix again by $\tilde{\mathbf{S}}$.

\subsubsection{Measurements with random configurations of individual loads}

The outcome of step 1 will be $\tilde{\mathbf{S}}$ with row-wise and column-wise scaling ambiguities for rows and lines associated with NDA ports, similar to Eq.~(\ref{eq_2555}). For $M_1$ random but known configurations of individual loads terminating the NDA ports, 
    \begin{equation}
        \mathbf{S}_{\mathrm{L},j} = \mathrm{diag} \left( \left[ r_{1,j}, r_{2,j}, \dots, r_{N_\mathrm{S},j} \right] \right),
    \end{equation}
where each $r_{i,j}$ is randomly chosen with equal probability from the set of the three available individual loads at the $i$th NDA port: $\{r_i^\mathrm{A},r_i^\mathrm{B},r_i^\mathrm{C}\}$, we measure the corresponding measurable scattering matrix $\hat{\mathbf{S}}_j$ which is defined by Eq.~(\ref{eq_load_S}). Based on our $M_1$ measurements, we generate $M_1-1$ triplets $\{\Delta\hat{\mathbf{S}}_j,\mathbf{S}_{\mathrm{L},j+1},\mathbf{S}_{\mathrm{L},j}  \}$, where 
\begin{equation}
\begin{split}
    \Delta\hat{\mathbf{S}}_j &= \hat{\mathbf{S}}_{j+1}-\hat{\mathbf{S}}_j \\ &= \mathbf{S}_\mathcal{AS} \left[ \left(  \mathbf{S}_{\mathrm{L},j+1}^{-1} -  \mathbf{S}_\mathcal{SS} \right)^{-1} - \left(  \mathbf{S}_{\mathrm{L},j}^{-1} -  \mathbf{S}_\mathcal{SS} \right)^{-1}  \right]  \mathbf{S}_\mathcal{SA}.
    \end{split}
\end{equation}
We then consider all entries of $\mathbf{S}_\mathcal{AS}$, $\mathbf{S}_\mathcal{SS}$ and $\mathbf{S}_\mathcal{SA}$ as independent complex-valued variables to be determined by gradient descent by minimizing the average magnitude of the entries of the difference between the predicted and measured $\Delta\hat{\mathbf{S}}_j$; the average is taken over all entries of $\Delta\hat{\mathbf{S}}_j$ and all realizations of $\Delta\hat{\mathbf{S}}_j$. This gradient-descent procedure is analogous to the corresponding part in~\cite{del2024virtual} except for the fact that we treat $\mathbf{S}_\mathcal{AS}$ and $\mathbf{S}_\mathcal{SA}$ as independent here, whereas they were the transpose of each other in~\cite{del2024virtual} concerned with reciprocal DUTs. Technical details on the gradient-descent algorithm can be found in Appendix~C of~\cite{del2024virtual}.
The outcomes are our estimates $\tilde{\mathbf{S}}_\mathcal{AS}$, $\tilde{\mathbf{S}}_\mathcal{SS}$ and $\tilde{\mathbf{S}}_\mathcal{SA}$ which are subject to the aforementioned row-wise and column-wise scaling ambiguities.

Then, we retrieve our estimate $\tilde{\mathbf{S}}_\mathcal{AA}$ in closed form (as we did in~\cite{del2024virtual}):
\begin{equation}
    \tilde{\mathbf{S}}_\mathcal{AA} \triangleq \Big\langle \hat{\mathbf{S}}_j - \tilde{\mathbf{S}}_\mathcal{AS}  \left(  \mathbf{S}_{\mathrm{L},j+1}^{-1} -  \tilde{\mathbf{S}}_\mathcal{SS} \right)^{-1}  \tilde{\mathbf{S}}_\mathcal{SA}\Big\rangle_j.
\end{equation}

At this stage, $\tilde{\mathbf{S}}$ is subject to row-wise and column-wise scaling ambiguities similar to those seen in Eq.~(\ref{eq_2555}).

\subsubsection{Measurements with random configurations of individual loads except for a 2PLN connecting an accessible and an NDA port}

The goal of step 2 is to identify $\epsilon_1$ and thereby lift the scaling ambiguity on the row and column of $\tilde{\mathbf{S}}$ associated with the first NDA port. To that end, we connect the last accessible port and the first NDA port to an arbitrary known 2PLN (as we did in step 5 of Sec.~\ref{step5}) and measure the measurable scattering matrix at the remaining $N_\mathrm{A}-1$ accessible ports for $M_2$ random known configurations of individual loads terminating the other $N_\mathrm{S}-1$ NDA ports.

This time, we conduct a gradient-descent with a single complex-valued variable to be determined by gradient descent: $\epsilon_1$. Specifically, we define
\begin{equation}
\tilde{\mathbf{S}}^\prime_{ij} =
\begin{cases}
\frac{1}{\epsilon_1^\prime}  \tilde{\mathbf{S}}_{ij}, & \text{if } i = \mathcal{S}_1,\, j \neq \mathcal{S}_1,\\[6pt]
\epsilon_1^\prime \tilde{\mathbf{S}}_{ij}, & \text{if } j = \mathcal{S}_1,\, i \neq \mathcal{S}_1,\\[6pt]
\tilde{\mathbf{S}}_{ij}, & \text{otherwise}.
\end{cases}
\end{equation}
and 
\begin{equation}\label{eq_3777}
    \hat{\mathbf{S}}_{M_1+j} = \tilde{\mathbf{S}}^\prime_{\mathcal{A}^\prime\mathcal{A}^\prime} +  \tilde{\mathbf{S}}^\prime_{\mathcal{A}^\prime\mathcal{S}^\prime} \left( \mathbf{S}_{\mathrm{L},M_1+j}^{-1} - \tilde{\mathbf{S}}^\prime_{\mathcal{S}^\prime\mathcal{S}^\prime} \right)^{-1} \tilde{\mathbf{S}}^\prime_{\mathcal{S}^\prime\mathcal{A}^\prime},
\end{equation}
where $\mathcal{A}^\prime = \mathcal{A}\setminus\mathcal{A}_{N_\mathrm{A}}$ (as before) and $\mathcal{S}^\prime = \mathcal{S}\cup\mathcal{A}_{N_\mathrm{A}}$.
Then, we perform a gradient descent to optimize our only variable $\epsilon_1^\prime$ such that the average magnitude of the entries of the difference between prediction and measurement of $\hat{\mathbf{S}}_{M_1+j} $ is minimized.
After convergence of the gradient descent, we use $\hat{\mathbf{S}}^\prime$ as our new $\hat{\mathbf{S}}$:
\begin{equation}
    \hat{\mathbf{S}} \leftarrow \hat{\mathbf{S}}^\prime.
\end{equation}
As evident from Eq.~(\ref{eq_3777}), only the row and column associated with the first NDA port are updated in this step 2; their scaling ambiguities are lifted.

\subsubsection{Measurements with random configurations of individual loads except for a 2PLN connecting two NDA ports}

The goal of step 3 is to lift the remaining ambiguities by identifying the remaining $\epsilon_i$ for $1<i\leq N_\mathrm{S}$ that are so far unknown. The procedure resembles that in step 2, and the general idea of identifying one unknown $\epsilon_i$ after the other by connecting a 2PLN to the $i$th and $j$th NDA ports, where $j$ is chosen such that $\epsilon_j$ is already known, resembles  step 6 of the closed-form method. Initially, only $\epsilon_1$ is already known. We connect a 2PLN to the first and second NDA ports, and measure the measurable scattering matrix at all $N_\mathrm{A}$ accessible ports for $M_2$ random configurations of individual loads terminating the other $N_\mathrm{S}-2$ NDA ports. Using the same gradient-descent procedure as in step 2, we identify $\epsilon_2$ and update $\tilde{\mathbf{S}}$ accordingly. Then, we connect a 2PLN to the second and third NDA ports to identify $\epsilon_3$, and so forth. Altogether, step 3 involves $(N_\mathrm{S}-1)M_2$ measurements and yields an ambiguity-free $\hat{\mathbf{S}}$ that approximates $\mathbf{S}$ -- provided that $M_1$ and $M_2$ were chosen ``sufficiently large'' for the given DUT and noise level; we are to date not aware of mathematically rigorous statements that could make the meaning of ``sufficiently large'' more concrete.

\section{Experimental Validation}\label{sec_exp_validation}

In this section, we experimentally validate our closed-form method from Sec.~\ref{subsec_closed_form} and our gradient-descent method from Sec.~\ref{subsec_grad_desc} considering an eight-port \textit{non-reciprocal} transmission-line network as DUT with four accessible and four NDA ports.

\subsection{Experimental Setup}

\begin{figure*}
    \centering
    \includegraphics[width=\textwidth]{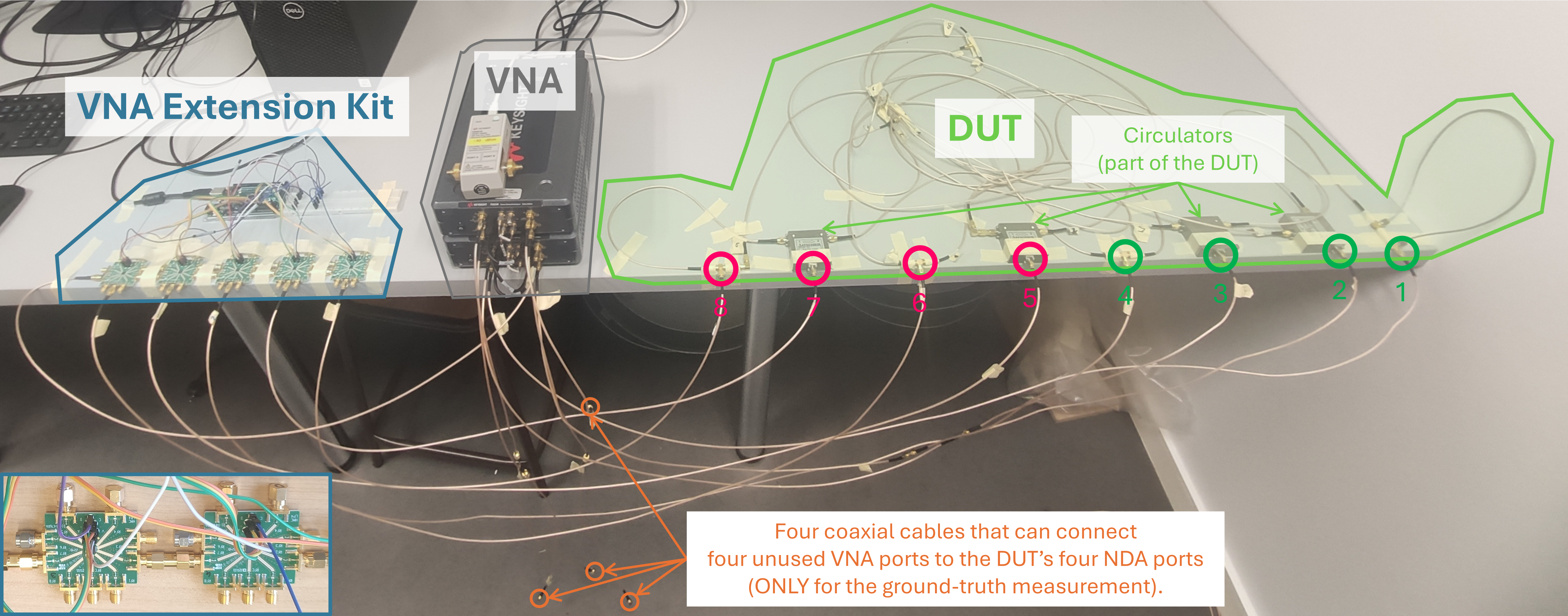}
    \caption{Experimental setup comprising an eight-port \textit{non-reciprocal} DUT (a complex transmission-line network involving four circulators), a four-port VNA, and our VNA Extension Kit with four ``virtual VNA ports''. The DUT's eight ports are partitioned into four accessible ports (1-4, green) and four NDA ports (5-8, magenta). The VNA Extension Kit comprises four switches (each connected to one of the DUT's four NDA ports) as well as an additional switch (the leftmost one) connected to the DUT's last accessible port, in line with the schematic in Fig.~\ref{Fig1}A. The VNA has four additional ports which are only used to measure the ground-truth DUT scattering matrix to validate our results. }
    \label{Fig3}
\end{figure*}

\subsubsection{Non-reciprocal DUT} The DUT is a complex transmission-line network\footnote{Alternative descriptions of our DUT as a non-reciprocal cable-network metamaterial or a non-reciprocal quantum-graph analogue are possible~\cite{sol2024covert}.} whose non-reciprocity originates from four circulators (C-C470.700-A1-(F1F1F1), Aerocomm). The DUT's non-reciprocity is apparent upon visual inspection of Fig.~\ref{Fig5}: The DUT's ground-truth scattering matrix (as well as our estimates thereof) is seen to be clearly non-symmetric.
The DUT is linear, passive and time-invariant. 
The DUT has eight monomodal ports with SubMiniature version A (SMA) connectors which are identified on the right side in Fig.~\ref{Fig3}; four of the DUT ports are treated as accessible (green) and the remaining four DUT ports are treated as NDA (magenta). Throughout this work, no DUT-specific a priori knowledge is used.

\subsubsection{VNA Extension Kit} 
\begin{figure}
    \centering
    \includegraphics[width=0.9\columnwidth]{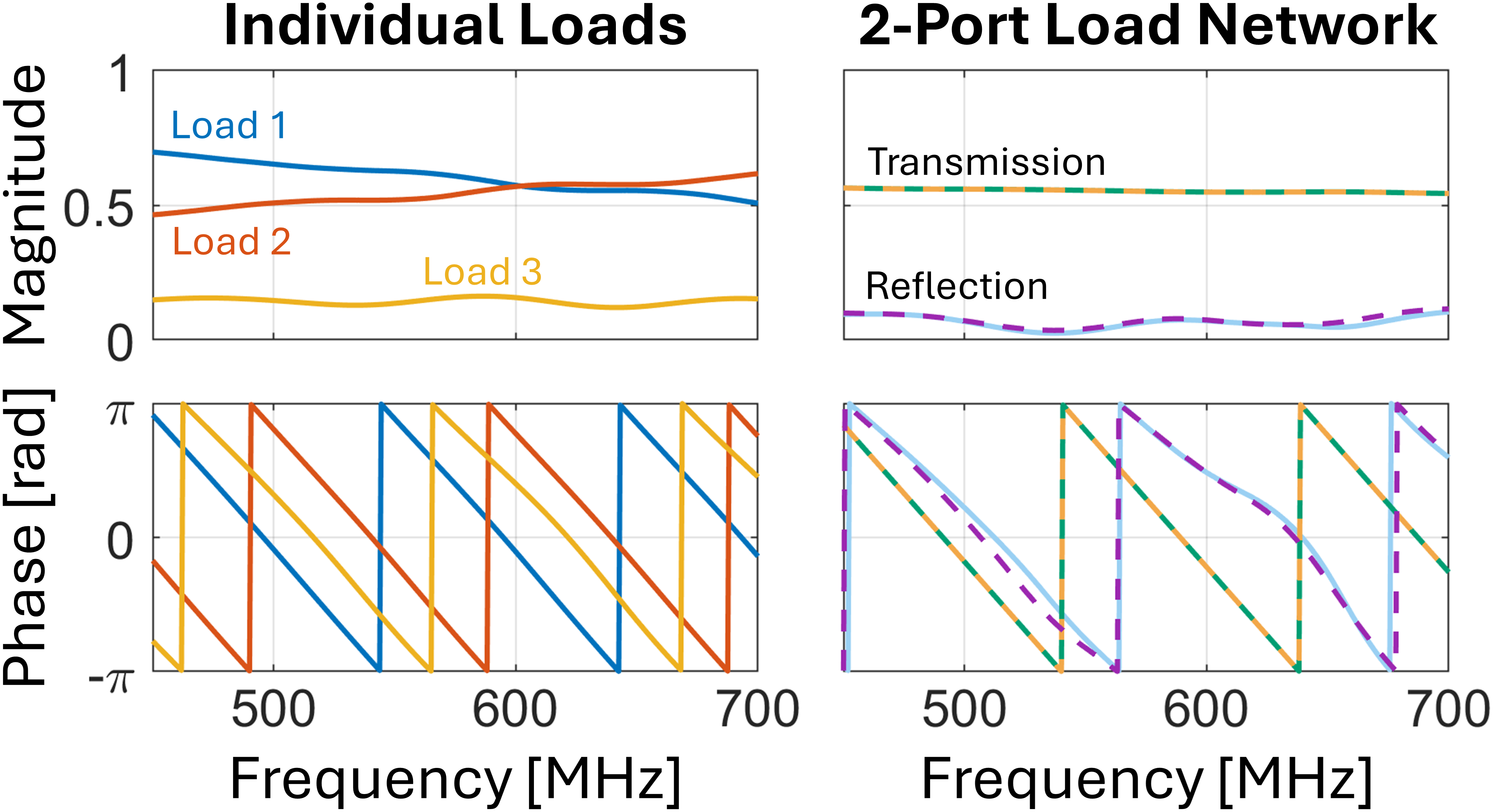}
    \caption{Measured scattering characteristics of the VNA Extension Kit (depicted in Fig.~\ref{Fig3} and inset therein). Magnitude and phase of the three individual loads are shown in the left column. Magnitude and phase of the four scattering coefficients of the 2PLN are shown in the right column.}
    \label{Fig4}
\end{figure}

Our VNA Extension Kit depicted on the left side of Fig.~\ref{Fig3} implements the schematic thereof seen in Fig.~\ref{Fig1}A based on GaAs MMIC SP8T switches (HMC321ALP4E, Analog Devices). Each switch has one SMA input port (to be connected via an SMA coaxial cable to an NDA port of the DUT) that can be connected to five distinct SMA output ports (three additional output ports would be available on the utilized SP8T switches but remain unused). As seen in the inset in Fig.~\ref{Fig3}, three of these output ports are terminated by standard SMA loads (open circuit, short circuit, and matched load) and the remaining two ports are connected via SMA connectors to neighboring switches. The corresponding measured scattering characteristics are displayed in Fig.~\ref{Fig4}. Importantly, we clearly see in Fig.~\ref{Fig4} that none of the three realized individual loads emulates a calibration standard (because of the wave propagation in the cables and switches). Besides four such switches, as seen in the schematic in Fig.~\ref{Fig1}A, one additional switch is required to enable switching between connecting the last accessible port to the VNA or via a 2PLN to the first NDA port. Thus, only two output ports of this additional switch are used.

\subsubsection{VNA} We conduct all measurements with an eight-port VNA (Keysight
M9005A) for 401 linearly spaced frequency points between 450 and 700~MHz (IFBW:~500~Hz, power:~12~dBm). We only use all eight VNA ports for the ground-truth measurement to validate our results. For the main measurements, only four VNA ports are used, while the other four VNA ports remain unused (see the identification of four coaxial cables connected to the unused VNA ports in Fig.~\ref{Fig3}).

\subsection{Measurement setup and methodology}

Our main measurement setup is depicted in Fig.~\ref{Fig3} and corresponds to the schematic in Fig.~\ref{Fig1}A. Note that the leftmost switch of the VNA Extension Kit is the one that is connected to the last accessible port (DUT port indexed 4), the second switch from the left of the VNA Extension Kit is connected to the first NDA port (DUT port indexed 5), etc. 
Compared to the earlier embodiment of a similar VNA Extension Kit as part of the experimental validation of the ``Virtual VNA 2.0'' for reciprocal DUTs in~\cite{del2024virtual2p0}, the present VNA Extension Kit \textit{fully} automates the switching between the required load configurations at the ``virtual VNA ports'', eliminating the need for any manual reconnections that would be tedious, less accurate, and error-prone. Thus, the use of the ``Virtual VNA'' with four actual and four virtual ports operationally resembles the use of a conventional VNA with eight actual ports in that in both cases only one manual connection between each DUT port and the corresponding VNA or Virtual VNA port needs to be established. 
The required load configurations for the two considered methods are summarized visually in Fig.~\ref{Fig1}B. For our experiment with  $N_\mathrm{A}=N_\mathrm{S}=4$, $M=19$ measurements are required for the closed-form  method from Sec.~\ref{subsec_closed_form}, and we choose $M_1 = 10 M_2 = 1000$ for the gradient-descent method from Sec.~\ref{subsec_grad_desc}.

Our measurement plane is at the DUT ports. The cables from VNA and VNA Extension Kit to the DUT are hence included in the VNA calibration and treated as part of the VNA Extension Kit, respectively. Note that the connection from the VNA to the last accessible DUT port (indexed 4) passes via the leftmost switch of the VNA Extension Kit (as seen in Fig.~\ref{Fig3} and the corresponding schematic in Fig.~\ref{Fig1}A), all of which is included in the VNA calibration.

\subsection{Experimental Results}

\begin{figure*}
    \centering
    \includegraphics[width=\textwidth]{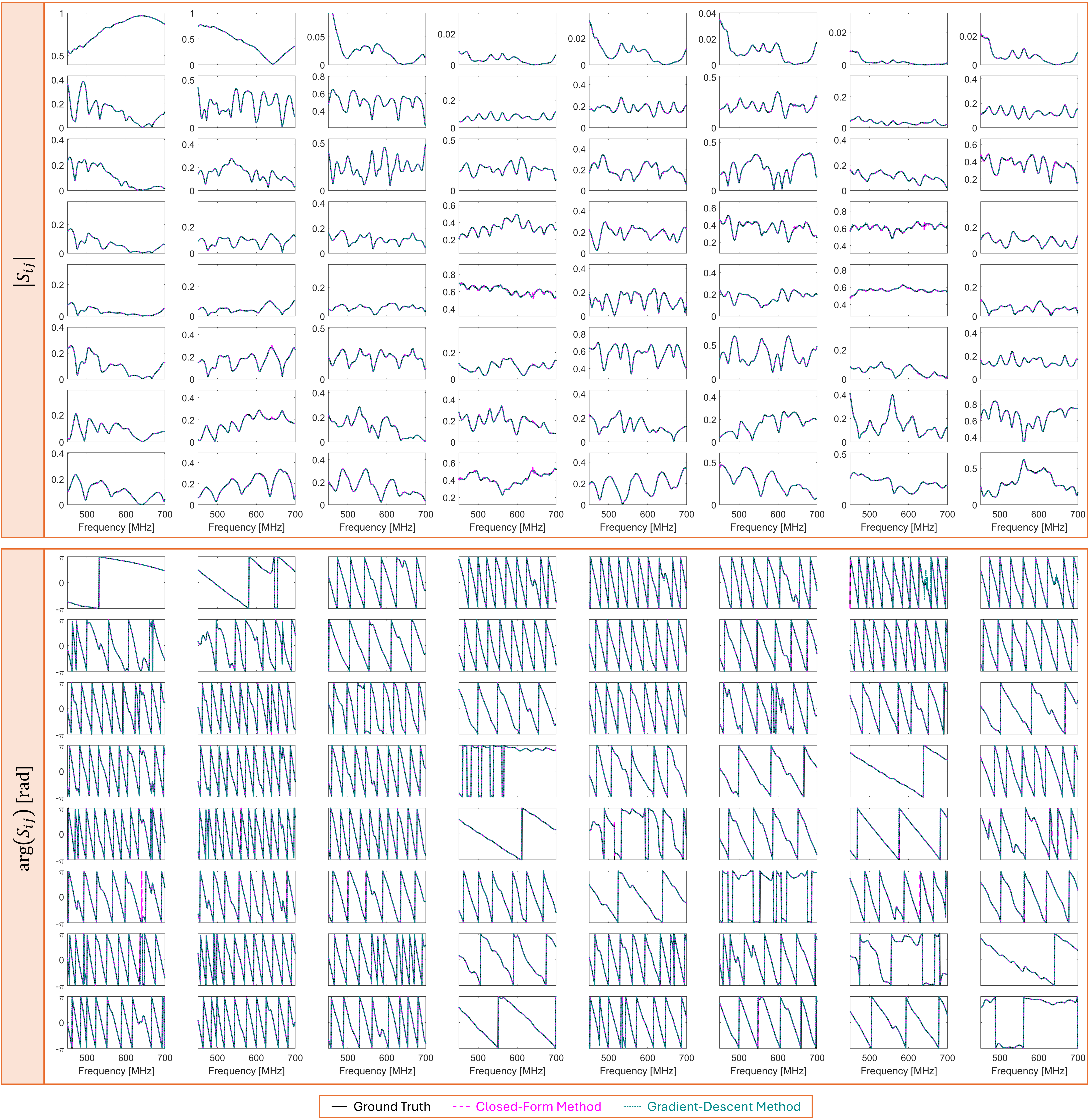}
    \caption{Comparison of ground truth (black continuous line), closed-form estimate (magenta dashed line) and gradient-descent estimate (dark cyan dotted line) for the \textit{non-reciprocal} DUT's $8\times8$ scattering spectra, in terms of their magnitudes (top panel) and phases (bottom panel). }
    \label{Fig5}
\end{figure*}

A comparison of the non-reciprocal DUT's ground-truth scattering matrix with our estimates thereof based on the closed-form method from Sec.~\ref{subsec_closed_form} and the gradient-descent method from Sec.~\ref{subsec_grad_desc} is displayed in Fig.~\ref{Fig5}. Upon visual inspection, the estimates appear to closely match the ground truth. Very minor deviations of the estimates from the ground truth can be spotted, e.g., for the magnitudes of $S_{47}$, $S_{54}$ and $S_{84}$ in the closed-form estimate at the 640~MHz frequency point and for the phase of $S_{17}$ in the gradient-descent estimate around 642~MHz where $|S_{17}|\approx0$. 

\begin{figure*}
    \centering
    \includegraphics[width=\textwidth]{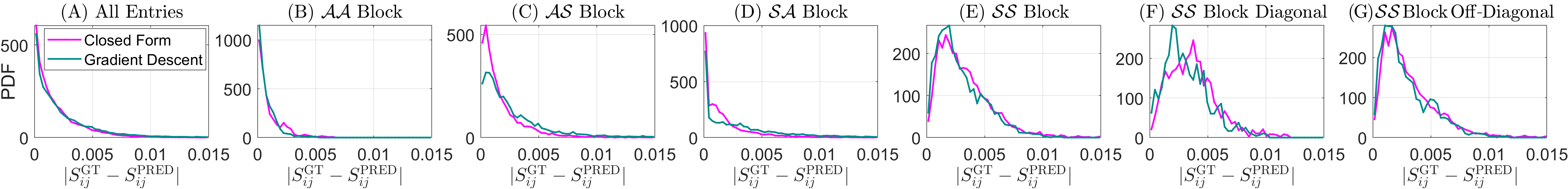}
    \caption{Probability density function (PDF) of the absolute error of the  scattering coefficients reconstructed with the two proposed methods: closed form (magenta) and gradient descent (dark cyan). The PDFs are evaluated across all 401 frequency points and additionally across all scattering coefficients (A) or only the scattering coefficients in the block $\mathcal{AA}$ (B), $\mathcal{AS}$ (C), $\mathcal{SA}$ (D) or $\mathcal{SS}$ (E), or only the diagonal entries of the block $\mathcal{SS}$ (F), or only the off-diagonal entries of the block $\mathcal{SS}$ (G).    }
    \label{Fig6}
\end{figure*}

To assess the accuracy of the two methods more quantitatively, we have evaluated the probability density functions (PDFs) of the absolute errors of the scattering coefficient estimates, which are shown in Fig.~\ref{Fig6}. The PDFs are based on all 401 frequency points and all scattering coefficients in Fig.~\ref{Fig6}A, as well as broken down to sub-groups of the scattering coefficients in Fig.~\ref{Fig6}B-G. In line with our observations from Fig.~\ref{Fig5}, no substantial accuracy differences between the two methods are discernible in Fig.~\ref{Fig6}. The shapes of the distributions are similar for both methods, and the distributions have a pronounced positive skew. The errors of the estimates of scattering coefficients from the $\mathcal{AA}$ block (associated with the DUT's accessible ports) are lowest.

Finally, to quantify the accuracy with a single metric, we evaluate
\begin{equation}
    \zeta = \left\langle\frac{\mathrm{SD}\left[S_{ij}^\mathrm{GT}(f)\right]}{\mathrm{SD}\left[S_{ij}^\mathrm{GT}(f) - S_{ij}^\mathrm{PRED}(f)\right]}\right\rangle_{i,j,f},
    \label{eq_zeta}
\end{equation}
where $\mathrm{SD}$ denotes the standard deviation, and the superscripts GT and PRED denote ground truth and prediction, respectively. Note that $\zeta$ is defined similar to a signal-to-noise ratio, treating the error as ``noise''.
A summary of this metric for the different considered groups of scattering coefficients is provided in Table~\ref{table_zeta}. The achieved accuracies are notably higher than in our previous experimental validation of the ``Virtual VNA 2.0''~\cite{del2024virtual2p0}, which we attribute to the fact that in the present experiment the DUT does not absorb waves as strongly as the DUT based on a reverberation chamber in~\cite{del2024virtual2p0}, as well as to the fact that the load switching is fully automated in the present work which eliminates inaccuracies related to manual interventions. Meanwhile, the differences between the closed-form method and the gradient-descent method are minute and not necessarily significant\footnote{We performed first the ground-truth measurement, second the $M$ measurements for the closed-form method, and third the $M_1+N_SM_2$ measurements for the gradient-descent method. Hence, the slightly worse performance of the gradient-descent method could be related to thermal drift over time or similar phenomena.}.

\begin{table}[htbp]
\centering
\caption{Performance of closed-form and gradient-descent methods in terms of $\zeta$ metric for different groups of scattering coefficients.}
\begin{tabular}{ |p{3cm}||>{\centering\arraybackslash}p{1.2cm}|>{\centering\arraybackslash}p{1.2cm}| }
 \hline
  & {Closed Form} & {Gradient Descent} \\
 \hline\hline
 {All Entries}       & 39.0~dB    & 37.0~dB \\
 {\(\mathcal{AA}\) Block}         & 46.3~dB    & 48.8~dB \\
 \(\mathcal{AS}\) Block     & 37.5~dB    & 34.5~dB \\
 \(\mathcal{SA}\) Block     & 37.0~dB    & 33.0~dB \\
 \(\mathcal{SS}\) Block     & 38.1~dB    & 39.2~dB \\
 \(\mathcal{SS}\) Block Diagonal     & 34.0~dB    & 35.3~dB \\
 \(\mathcal{SS}\) Block Off-Diagonal & 38.7~dB  & 39.8~dB     \\
 \hline
\end{tabular}
\label{table_zeta}
\end{table}

\section{Conclusion}\label{sec_conclusion}

To summarize, we have generalized the ``Virtual VNA'' concept to non-reciprocal DUTs. Thereby, we enable the unambiguous estimation of a non-reciprocal $N$-port DUT's scattering matrix using a VNA with $1<N_\mathrm{A} < N$ ports without any need for reconnections. The DUT's remaining $N_\mathrm{S}=N-N_\mathrm{A}$ NDA ports  are hence not connected to the VNA but to ``virtual VNA ports'' of a VNA Extension Kit. This kit terminates the DUT's NDA ports with  known loads that are tunable in accordance with certain requirements (at least three distinct individual loads and coupled loads are available). 
First, we derived a closed-form method that requires a specific set of load configurations. Second, we established a complimentary gradient-descent method that can be applied to an arbitrary number of arbitrary load configurations. Interestingly, the required VNA Extension Kit is the same as for reciprocal DUTs in the ``Virtual VNA 2.0'' technique. However, each coupled load is required to identify a complex-valued scalar in the case of non-reciprocal DUTs, which specializes to only choosing a sign in the case of reciprocal DUTs~\cite{del2024virtual2p0}. 
We validated both ``Virtual VNA 3.0'' methods experimentally considering a non-reciprocal transmission-line network with $N_\mathrm{A}=N_\mathrm{S}=4$. Both methods achieved comparable high accuracies. The closed-form method did so based on fewer measurements, suggesting that it is preferable under the low-noise measurement conditions of the present work. A systematic exploration of the performance of both methods as a function of measurement noise, as well as on the chosen number of measurements $M_1$ and $M_2$ in the gradient-descent method, is left for future work. Similar to corresponding results for reciprocal DUTs in~\cite{del2024virtual}, we expect the gradient-descent method to be advantageous under noisier measurement conditions.

The two developed Virtual VNA methods for \textit{non-reciprocal} DUTs can also be applied to reciprocal DUTs; however, if the DUT is known to be reciprocal, it is preferable to constrain the estimation problem by accounting for the knowledge that the DUT's scattering matrix is symmetric. Our previous works on the Virtual VNA for reciprocal DUTs~\cite{del2024virtual,del2024virtual2p0} showed that the gradient-descent method can also be successfully applied to non-coherent measurements. A similar conclusion does \textit{not} apply to the present case of non-reciprocal DUTs because ambiguities about the relative phases between different rows of $\mathbf{S}$ cannot be resolved. However, we expect that tunable interferences of the outgoing wavefronts can resolve these ambiguities and we will study this in the future.\footnote{A conceptually similar method was used to resolve row-wise ambiguities in the phase retrieval of an off-diagonal block of a reciprocal DUT's scattering matrix in~\cite{goel2023referenceless}.}

\appendix

\label{Appendix_redheffer}
The (\textit{forward}) Redheffer star product $    {\mS}^{\U\V} = \mS^\U \underset{\iC_\U,\iC_\V}{\star} \mS^\V$ from Eq.~(\ref{eq:redheffer_star}) is defined as 
\begin{equation}
\begin{split}
& {\mS}^{\U\V}_{\iN_\U\iN_\U} = \mS^\U_{\iN_\U\iN_\U} - \mS^\U_{\iN_\U\iC_\U} \mS^\V_{\iC_\V\iC_\V} \mathbf{X}^{\U\V} \mS^\U_{\iC_\U\iN_\U},\\
& {\mS}^{\U\V}_{\iN_\U\iN_\V} = -\mS^\U_{\iN_\U\iC_\U} \mathbf{X}^{\V\U} \mS^\V_{\iC_\V\iN_\V},\\
& {\mS}^{\U\V}_{\iN_\V\iN_\U} = -\mS^\V_{\iN_\V\iC_\V} \mathbf{X}^{\U\V} \mS^\U_{\iC_\U\iN_\U},\\
& {\mS}^{\U\V}_{\iN_\V\iN_\V} = \mS^\V_{\iN_\V\iN_\V} - \mS^\V_{\iN_\V\iC_\V} \mS^\U_{\iC_\U\iC_\U} \mathbf{X}^{\V\U} \mS^\V_{\iC_\V\iN_\V},
\end{split}
\label{eq:redheffer}
\end{equation}
where 
\begin{equation}
\begin{split}
 & \mathbf{X}^{\U\V} = \left( \mS^\U_{\iC_\U\iC_\U} \mS^\V_{\iC_\V\iC_\V} - \mI_{N_\mathrm{C}} \right)^{-1}, \\
& \, \mathbf{X}^{\V\U} = \left( \mS^\V_{\iC_\V\iC_\V} \mS^\U_{\iC_\U\iC_\U} - \mI_{N_\mathrm{C}} \right)^{-1},
\end{split}
\label{eq:redheffer_aux}
\end{equation}
and $\mI_{N_\mathrm{C}}$ denotes the $N_\mathrm{C} \times N_\mathrm{C}$ identity matrix.

The \textit{inverse} Redheffer star product ${\mS}^{\U} = \mS^{\U\V} \overset{-1}{\underset{\iC_\U,\iC_\V}{\star}} \mS^\V$ from Eq.~(\ref{eq:inverse_redheffer_star}) is implemented by first defining an auxiliary matrix
\begin{equation}
    \mathbf{R} 
    \;=\; 
    \bigl[\mS^\V_{\iN_\V\,\iC_\V}\bigr]^{-1}
    \;\Bigl(\mS^\V_{\iN_\V\,\iN_\V} \;-\; \mS^{\U\V}_{\iC_\U\,\iC_\U}\Bigr)\;
    \bigl[\mS^\V_{\iC_\V\,\iN_\V}\bigr]^{-1}.
\end{equation}
We then solve for the $\iC_\U\,\iC_\U$ block of \(\mS^\U\) via
\begin{equation}
    \mS^\U_{\iC_\U\,\iC_\U}
    \;=\;
    -\,\Bigl(\mI_{N_\mathrm{C}} - \mathbf{R}\,\mS^\V_{\iC_\V\,\iC_\V}\Bigr)^{-1}\,\mathbf{R}.
\label{eq:s_cc_inverse}
\end{equation}
Once \(\mS^\U_{\iC_\U\,\iC_\U}\) is determined, we evaluate $\mathbf{X}^{\U\V} $ and $\mathbf{X}^{\V\U} $ using Eq.~(\ref{eq:redheffer_aux}). The remaining blocks of
\(\mS^\U\) then follow from
\begin{equation}
\begin{split}
& \mS^\U_{\iN_\U\,\iC_\U}
\;=\;
-\,\mS^{\U\V}_{\iN_\U\,\iC_\U}
\;
\bigl[\mathbf{X}^{\V\U}\;\mS^\V_{\iC_\V\,\iN_\V}\bigr]^{-1},
\\[6pt]
& \mS^\U_{\iC_\U\,\iN_\U}
\;=\;
-\,\bigl[\mS^\V_{\iN_\V\,\iC_\V}\;\mathbf{X}^{\U\V}\bigr]^{-1}
\;\mS^{\U\V}_{\iC_\U\,\iN_\U},
\\[6pt]
& \mS^\U_{\iN_\U\,\iN_\U}
\;=\;
\mS^{\U\V}_{\iN_\U\,\iN_\U}
\;+\;
\mS^\U_{\iN_\U\,\iC_\U}
\;\mS^\V_{\iC_\V\,\iC_\V}
\;\mathbf{X}^{\U\V}
\;\mS^\U_{\iC_\U\,\iN_\U}.
\end{split}
\label{eq:s_blocks_inverse}
\end{equation}

\bibliographystyle{IEEEtran}

\providecommand{\noopsort}[1]{}\providecommand{\singleletter}[1]{#1}%

\end{document}